\documentclass{article}
\usepackage{latexsym}
\usepackage{color,cite,graphicx}
\usepackage{latexsym,amssymb,epsf} 

\newcommand{\bge}{\begin{equation}}
\newcommand{\ee}{\end{equation}}
\newcommand{\bgc}{\begin{center}}
\newcommand{\ec}{\end{center}}
\newcommand{\bgea}{\begin{eqnarray}}
\newcommand{\eea}{\end{eqnarray}}
\newcommand{\bgeas}{\begin{eqnarray*}}
\newcommand{\eeas}{\end{eqnarray*}}

\begin{document}

\title{
  \bf Large scale molecular dynamics simulation of self-assembly
      processes in short and long chain cationic surfactants
}

\author{
  J.-B. Maillet$^1$, V. Lachet$^2$,\\
  {\footnotesize Schlumberger Cambridge Research, High Cross,
    Madingley Road,}\\
  {\footnotesize Cambridge, CB3 0EL, U.K.}\\
  Peter V. Coveney,\\
  {\footnotesize Centre for Computational Science,
    Department of Chemistry,}\\
  {\footnotesize Queen Mary and Westfield College,
    University of London, Mile End Road,}\\
  {\footnotesize London E1 4NS, U.K.}\\
  {\footnotesize{\tt p.v.coveney@qmw.ac.uk}}\\[1.5cm]
  {\footnotesize $^1$ Present address: CECAM, Ecole Normale
    Sup\'{e}rieure,}\\
  {\footnotesize 46, All\'ee d'Italie, 69364 Lyon Cedex 07,
  France.}\\[0.3cm] 
  {\footnotesize $^2$ Present address: Institut Fran\c{c}ais du
  P\'etrole, 1 et 4 avenue de Bois-Pr\'eau,}\\ 
  {\footnotesize 92852 Rueil-Malmaison, France.}\\[0.3cm]
}

\maketitle

\newpage

\begin{abstract}
We report on an investigation of the structural and dynamical properties
of \emph{n}-nonyltrimethylammonium chloride ($\rm C_{9}TAC$) and erucyl \emph{bis}
[2-hydroxyethyl] methylammonium chloride (EMAC) micelles in aqueous solution. A
fully atomistic description was used, and the time evolution was
computed using molecular dynamics. The calculations were
performed in collaboration with Silicon Graphics Inc. using the Large-scale
Atomic/Molecular Massively Parallel Simulator (LAMMPS) code~\cite{lammps} on
a range of massively parallel platforms. Simulations were
carried out in the isothermal-isobaric (N,P,T) ensemble, and run for
up to 3 nanoseconds. Simulated systems contained approximately 50
surfactant cations and chloride counterions, surrounded by 3000 water molecules. 
Starting from different initial configurations (spherical micelle,
wormlike micelle) in the case of the $\rm
C_{9}TAC$ molecule, we observe shape transformations on the
timescale of nanoseconds, micelle fragmentations, and
surfactant-monomer exchange with the
surrounding medium.
Starting from a random distribution of surfactant molecules in the solution, we
observe the mechanism of micelle formation at the
molecular level.
The mechanism of self-assembly or fragmentation of a micelle is
interpreted in terms of generalised classical nucleation theory.
Our results indicate that, when these systems are far from
equilibrium and at high surfactant concentration, the basic
aggregation-fragmentation mechanism is of Smoluchowski type
(cluster-cluster coalescence and break up); closer to equilibrium and
at lower surfactant concentration, this mechanism appears to follow a
Becker-D\"{o}ring process (stepwise addition or removal of
surfactant monomers).
In the case of the EMAC molecule, we have characterised two different
structures (spherical and cylindrical) of the micelle, and have found
that water penetration is not important. We have
also studied the effect
of the introduction of co-surfactant (salicylate)
molecules to the EMAC system; hydrogen bonds between surfactant head
groups and co-surfactant molecules were observed to play an important
role in stabilizing wormlike micelles.
\end{abstract}

\newpage


\section{Introduction}
For the past two decades, amphiphilic systems have constituted a field
of great interest, both from a fundamental and an industrial point of
view~\cite{gompper}. This is mainly due to the fact that they
exhibit a rich phase
diagram when mixed with water and other organic species. Their
amphiphilic nature leads to the formation of self-assembled mesoscopic
structures,
for example spherical micelles at intermediate surfactant concentrations. At
higher concentrations, or with the addition of electrolyte or organic
compounds (co-surfactant), cylindrical wormlike micelles may be
stabilized. These wormlike micelles confer visco-elastic
properties to the fluid. Indeed, gels are frequently formed,
reflecting the entanglement between worms, with viscosity dependent on
temperature, salt concentration, etc. As the surfactant
concentration increases, new structures appear, including vesicles and
bilayers. The rheology of these amphiphilic systems is often analysed
using a simple theory of self-assembly which combines
both thermodynamics and geometrical
arguments~\cite{israelachvili,israel2}. According to this approach, the
ability of
surfactant molecules to form a particular type of structure is
governed by the packing parameter $P$, which is the ratio of the volume
of the hydrophobic tail $v$ to the length of this tail $l$ times the
effective surface area per head group $a$:
\begin{equation}
P=\frac{v}{la}.
\end{equation}
Considering $v$ and $l$ to be
constant for a particular molecule, the geometry of the self-assembled
structure
is controlled by the effective surface area $a$ of the head group (for
example, the addition of
electrolyte tends to screen the repulsive interaction between head
groups, and so to decrease the effective head group area, leading to a
transition from spherical to wormlike micelles).\\
Molecular dynamics simulations have succeeded in resolving
the detailed structure of spherical micelles, and their interactions
with the solvent. These simulations have commonly involved 10 to 20
surfactant molecules with up to 1000 water molecules, and have been run for
a few hundred picoseconds~\cite{kuhn,laaksonen,kuhn2,watanabe,watanabe2}. Under these conditions, the dynamical
properties of micelles are extremely difficult to study because the
time scales of dynamical processes may vary from $10^{-11}$
(presumed characteristic time of shape fluctuations) to $10^{4}$ seconds (relaxation
time of some shear-induced structures).
The length and timescales involved in self-assembly phenomena
correspond typically to the domain of applicability for different
mesoscopic models including lattice-gas~\cite{boghosian,emerton,bcl},
lattice-Boltzmann~\cite{higuera,cbc} and dissipative particle dynamics
(DPD)~\cite{hoogerbrugge, boek2}. An
outstanding challenge for
these upscaling methods is the definition of a coherent link between the
atomistic description of the sytem and the related meso and macroscopic
behaviour. Some recent work~\cite{flekkoy,fc} has shown the feasibility
of such an approach based on a systematic coarse graining of the system in
order to link DPD to the microscopic world.
Smit \emph{et al.}~\cite{smit} earlier developed a molecular dynamics
model using a simplified description of the component molecules.
This allows a prediction of the
size distribution of micelles which agrees qualitatively with both experiments
and theory. Nevertheless, chemical specificity is not taken into
account in this model.
In this paper, we report a fully atomistic study of the structure and
dynamics of micelles, based on trajectories calculated up to 3 nanoseconds.
On this timescale, we can study some fast dynamical processes, such as
monomer insertion or removal from a micelle as well as local growth and
fragmentation for individual micelles, which occur much faster than the
approach to global thermodynamic equilibrium. The kinetics of such processes
may be modelled on the basis of a generalisation of
Classical Nucleation Theory (CNT)~\cite{doring,volmer1,volmer2,volmer3}. This theory
considers clusters of different sizes, which exchange monomers with the
surrounding medium (solvent + surfactant monomers), but neglects
cluster-cluster interactions. It provides a description of the time
evolution of the size distribution of clusters. A development
of this model including inhibition phenomena has recently been
published~\cite{wattis}.
The Becker-D\"{o}ring equations of CNT 
are a special case of the more general discrete
Smoluchowski coagulation-fragmentation equations which describe rate 
processes between two clusters of size $r$ and $s$ and a cluster of
size $r+s$. Although the determination of the rate coefficients for each
single process is not possible from a small number of molecular
dynamics trajectories, we are mainly interested here in analysing the
dominant
mechanism (Becker-D\"{o}ring or Smoluchowski) when systems are not at
equilibrium.
To the best of our knowledge, this work constitutes the first fully
atomistic approach to such phenomena on this timescale, for systems
containing up to 15000 atoms. This has been
achieved with the conjugate use of large parallel computers and a
highly scalable molecular dynamics program.

\section{Description of the model}
\subsection{Surfactant and water molecules}
\begin{figure}[p]
\begin{minipage}[t]{14.0cm}
\begin{minipage}[t]{7.0cm}
\centering
\scalebox{0.4}{\includegraphics{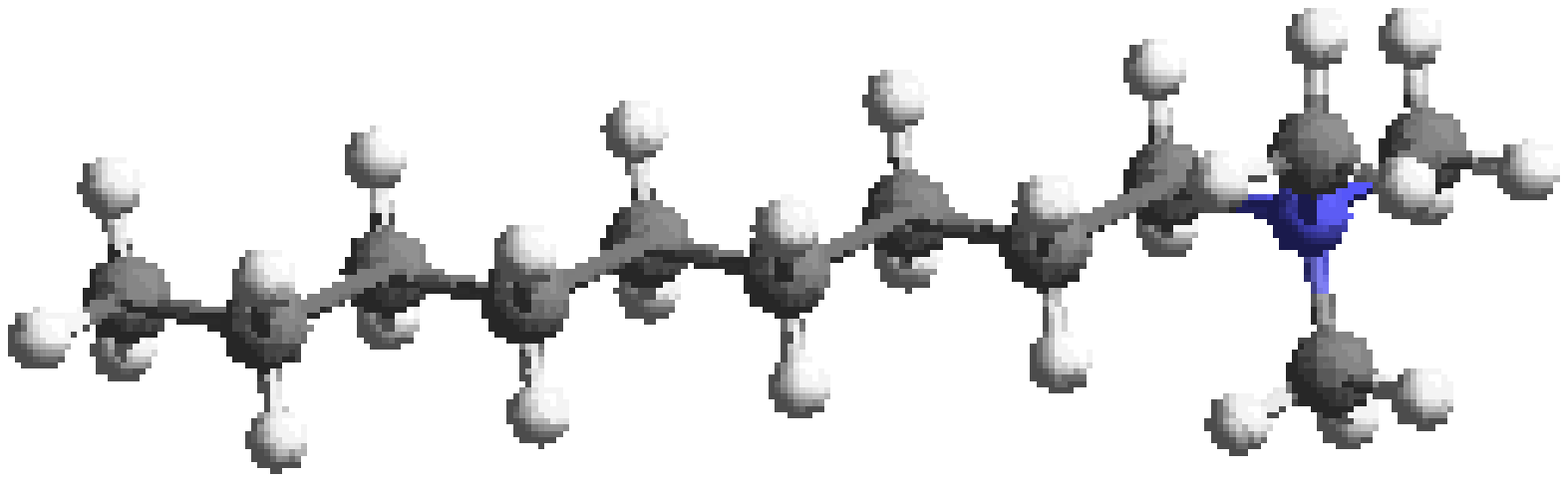}}
\end{minipage}
\hspace{2cm}\begin{minipage}[t]{7.0cm}
\scalebox{0.4}{\includegraphics{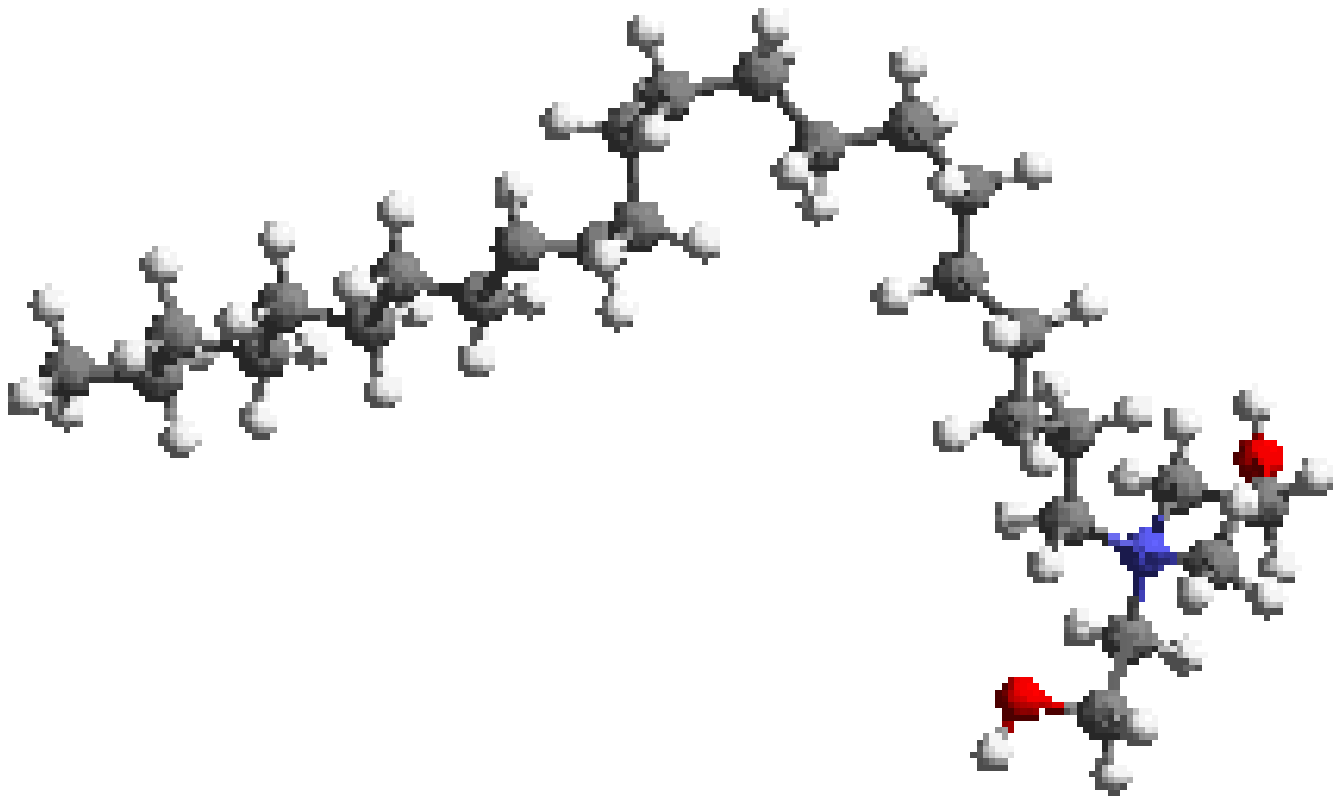}}
\end{minipage}
\end{minipage}
\caption{\textbf{Left: View of the $\rm \mathbf C_{9}TAC$ surfactant molecule.
Right: View of the EMAC molecule.}}
\label{molecule}
\end{figure}

Two surfactant molecules have been studied in aqueous solution, namely
\emph{n}-nonyltrimethylammonium chloride ($\rm C_{9}TAC$), and erucyl \emph{bis}
[2-hydroxyethyl] methylammonium chloride (EMAC). The latter is known
to form wormlike micellar viscoelastic fluids which find commercial
use in hydraulic fracturing operations~\cite{chase}.
The two surfactant molecules are shown
in figure~\ref{molecule}. In some simulations,
salicylate co-surfactant molecules have been added to the EMAC
solution.
The force-field
used to model these surfactant molecules is the ``constant
valence force-field'' (cvff)~\cite{hagler}, which includes explicitly
all the atoms. The total energy is the sum of the intramolecular and
intermolecular interactions. The intramolecular interactions are
represented as a sum of four types of term:
\begin{equation}
\varepsilon_{intra}=\sum \varepsilon_{stretch} + \sum
\varepsilon_{bend} + \sum \varepsilon_{torsion} + \sum \varepsilon_{out-of-plane} .
\label{somme}
\end{equation}
The explicit forms for each term in eq~(\ref{somme}) are given in eqs (\ref{stretch}-\ref{outofplane}). The
bond stretching term is given by:
\begin{equation}
\varepsilon_{stretch}=\sum_{i} k^b_{i}(b-b_{0})^{2},
\label{stretch}
\end{equation}
where $k^b_{i}$ is the bond stretching force constant, $b_{0}$ is the
equilibrium bond length, and $b$ is the actual bond length. The bond
bending term is given by eq~(\ref{bending}) :
\begin{equation}
\varepsilon_{bending}=\sum_{i} k^\theta_{i}(\theta-\theta_{0})^{2},
\label{bending}
\end{equation}
where $k^\theta_{i}$ is the bond bending force constant, $\theta_{0}$ is the
equilibrium bond angle, and $\theta$ is the actual bond angle. The
torsional contribution to the intramolecular energy is represented by
a cosine series:
\begin{equation}
\varepsilon_{torsion}=\sum_{i} k^\phi_{i}(1 + Scos(n \phi)),
\label{torsion}
\end{equation}
where $k^\phi_{i}$ is the torsional force constant, S is a phase factor
(equal to $1$ or $-1$ depending on the dihedral angle considered),
and $\phi$ is the torsional angle. The out-of-plane term describes the
resistance to out-of-plane bending and is expressed by a quadratic
distortion potential function:
\begin{equation}
\varepsilon_{out-of-plane}=\sum_{i} k^\chi_{i} \chi^{2},
\label{outofplane}
\end{equation}
where $k^\chi_{i}$ is the bending constant and $\chi$ is the bending angle.
The expression for the non-bonded interactions is:
\begin{equation}
\varepsilon_{inter}=\sum \varepsilon_{vdW} + \sum \varepsilon_{coulombic}
\label{nonbounded}
\end{equation}
where the summations are performed over all the
non-bonded pairs of atoms.
A Lennard-Jones 12-6 pair interaction is used for the van der Waals
energy, $\varepsilon_{vdW}$, and the partial charges involved in the
coulombic term were computed using the semi-empirical quantum
mechanical program MOPAC~\cite{mopac}, within the AM1 approximation.
The charge on the
$-N(CH_{3})_{3}$ head group of the $\rm C_{9}TAC$ surfactant was found to
be $+0.88$, and $+0.90$ for the EMAC head group
$-N(C_{2}H_{4}OH)_{2}CH_{3}$.
Water molecules are represented using the Jorgensen~\cite{jorgensen}
TIP3P model: interactions between water molecules are described by a
Lennard-Jones potential between oxygen atoms and electrostatic
contributions between all atoms (hydrogen and oxygen). All the
parameters of the TIP3P model are shown in table~\ref{waterpot}.
In order to check the performance of the TIP3P water force-field in
predicting the bulk properties of water, we performed a molecular
dynamics simulation of liquid water and compared the computed properties
such as water density, diffusion coefficient and radial
distribution function between oxygen atoms to the corresponding
experimental data. More
details about the simulation procedure can be found in
ref~\cite{boek}. The calculated bulk water density and self-diffusion
coefficient as averages over the stored time series of particle
coordinates are listed in table~\ref{waterprop}.
Both values compare well with experiments.
\subsection{Molecular dynamics method}
Several distinct initial configurations of surfactant molecules
surrounded by water molecules were constructed. They
consisted of an infinite wormlike micelle, a spherical micelle,
or a random distribution of surfactant molecules. Forty-eight or fifty
surfactant molecules were employed in the cases of
$\rm C_{9}TAC$ and EMAC respectively. The number of
surrounding water molecules was approximately equal to 3000 in each
simulation. In some cases, electrolyte (NaCl) and/or co-surfactant
(salicylate) was also added to the solution.\\
At the beginning of the dynamics simulation the total potential
energy was minimised in order to generate a reasonable starting point.
To carry out the minimisation, we used a truncated Newton-Raphson
method requiring evaluation of the second derivative of the
potential energy with respect to the atomic coordinates. After
minimisation, random velocities selected according to a Maxwellian distribution
at a temperature of 300 K were assigned to each atom. The pressure
was set to 1 atm and the temperature was fixed at 300 K. Pressure and
temperature were controlled using the Nos\'{e}-Hoover algorithm~\cite{nose}.
In most of our studies, the total simulation time was larger than 1
ns. To integrate the Newtonian equations of motion for all atoms, we
used the Verlet leapfrog algorithm with a timestep of 1 fs.
Periodic boundary conditions were applied in all three spatial
directions and Ewald summation was used to handle the long-range
electrostatic interactions in conjunction with the particle-particle /
particle mesh (PPPM) method, an $\mathcal{O}(NlogN)$ algorithm~\cite{frenkel}. A
cut-off radius of $10.0$ $\rm \AA$ was used for non-coulombic interactions.

\subsection{Parallel implementation of molecular dynamics}
\begin{figure}[p]
\scalebox{0.8}{\includegraphics{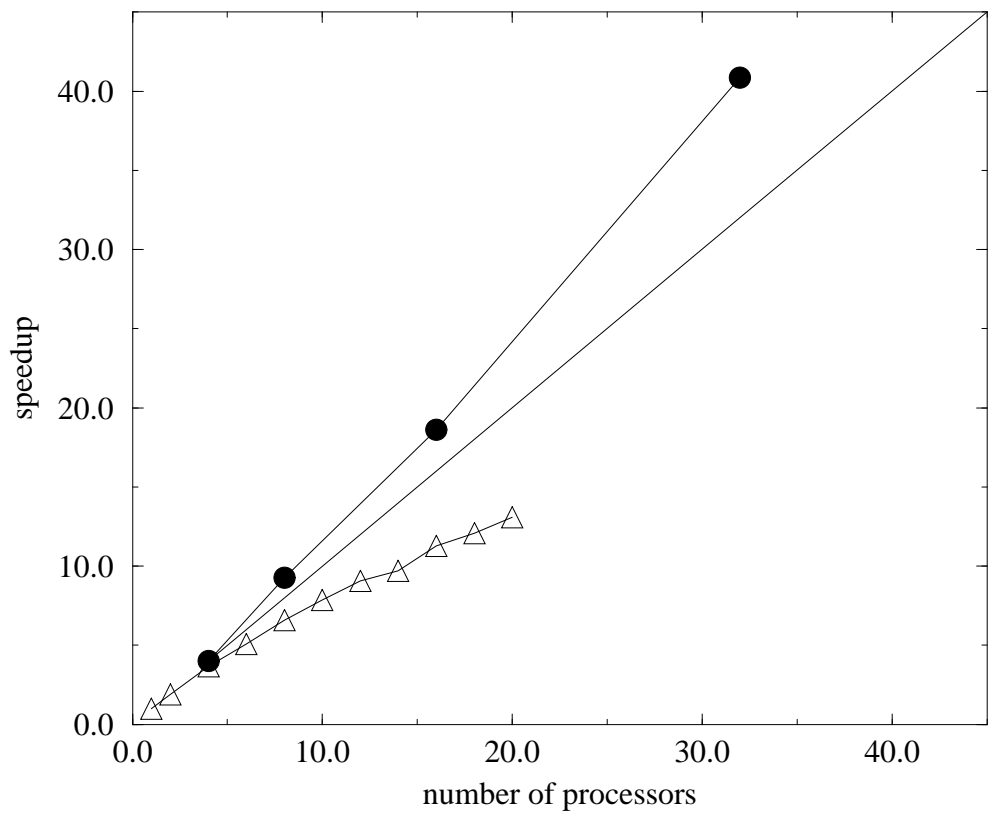}}
\caption{\textbf{Results of performance tests on a Silicon Graphics Origin 2000
for two different parallel molecular dynamics codes: LAMMPS (circles)
and $\mathbf \rm Cerius^2$ (triangles). The straight line represents
the theoretical optimum performance based on single processor speed,
as in eq~\ref{speedup}.}}
\label{bench}
\end{figure}

All MD simulations were carried out either on a Silicon Graphics Origin 2000
or on a Cray T3E using the Large-scale Atomic/Molecular Massively
Parallel Simulator (LAMMPS) code~\cite{lammps}. LAMMPS is a highly scalable
classical molecular dynamics code designed for simulating molecular
and atomic systems on parallel supercomputers. To study large systems
of molecules for a large number of time steps, an algorithm is required that has a
very good speedup with the number of processors used. This
speedup has been calculated up to 1024 processors on a 1500 node T3E, and the
results display the desired linear scalability property. The same
calculations have also been performed
on a Silicon Graphics Origin 2000 using up to 32 processors. Results of the
benchmarks are displayed in figure~\ref{bench}, where the speedup is
given by:
\begin{equation}
\rm speedup(k) = \frac{time(1 \:processor)}{time (k \:processors)}.
\label{speedup}
\end{equation}
One can see that the parallel
performance of the LAMMPS code is superior to that of the MD codes in
the commercial package
$\rm Cerius^2$~\cite{cerius}.
The superlinear behaviour observed in the case of
LAMMPS, related to the efficient calculation of the coulomb
interactions and the use of spatial domain decomposition,
can be understood in term of cache memory utilisation which is
sub-optimal for one (or a very small number) of processors.

\section{Structure and dynamics of $\rm \mathbf C_{9}TAC$ micelles}
\label{restac}
The instantaneous configurations of the system are displayed using the
MSI $\rm Cerius^{2}$ package~\cite{cerius}.
Some images employ Connolly surfaces~\cite{connoly} calculated specifically for
some molecules or fragments. A Connolly surface is the van der Waals
surface of the molecule/fragment that is accessible to a solvent
molecule. In all cases studied in this paper, the solvent molecule
is a water molecule. The blue square appearing in each snapshot
represents the simulation box.
\subsection{Spherical micelle}
\begin{figure}[p]
\begin{center}
\begin{tabular}{lr}
\includegraphics[scale=0.5]{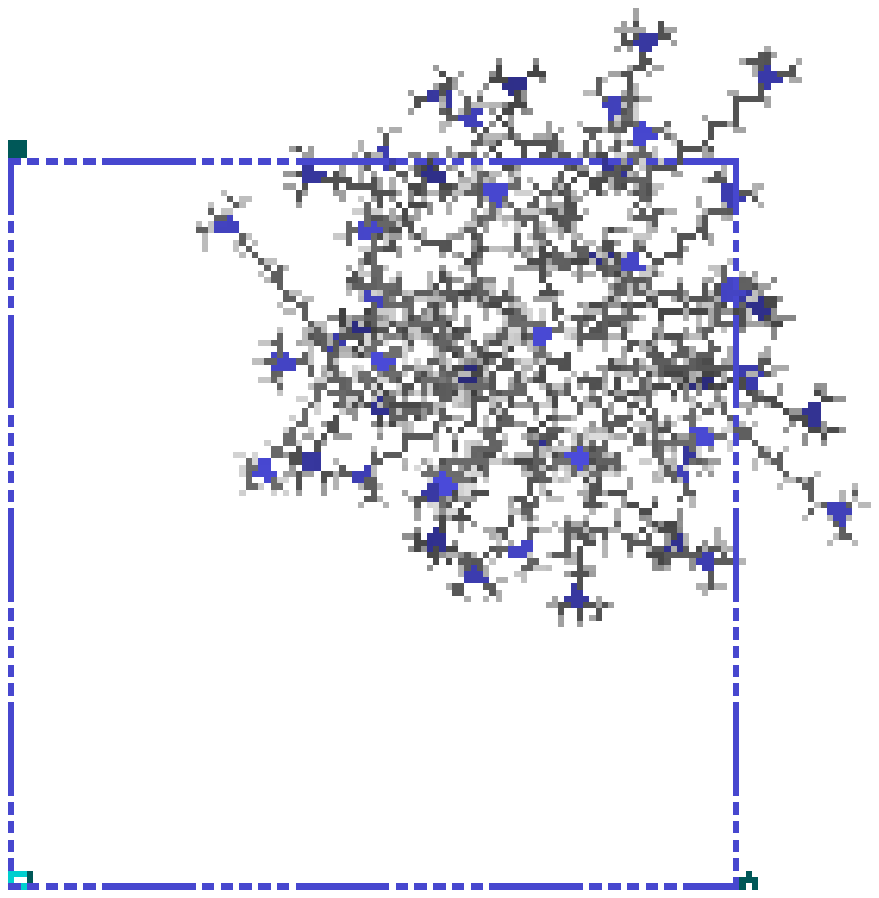}&
\hspace{2cm}\includegraphics[scale=0.5]{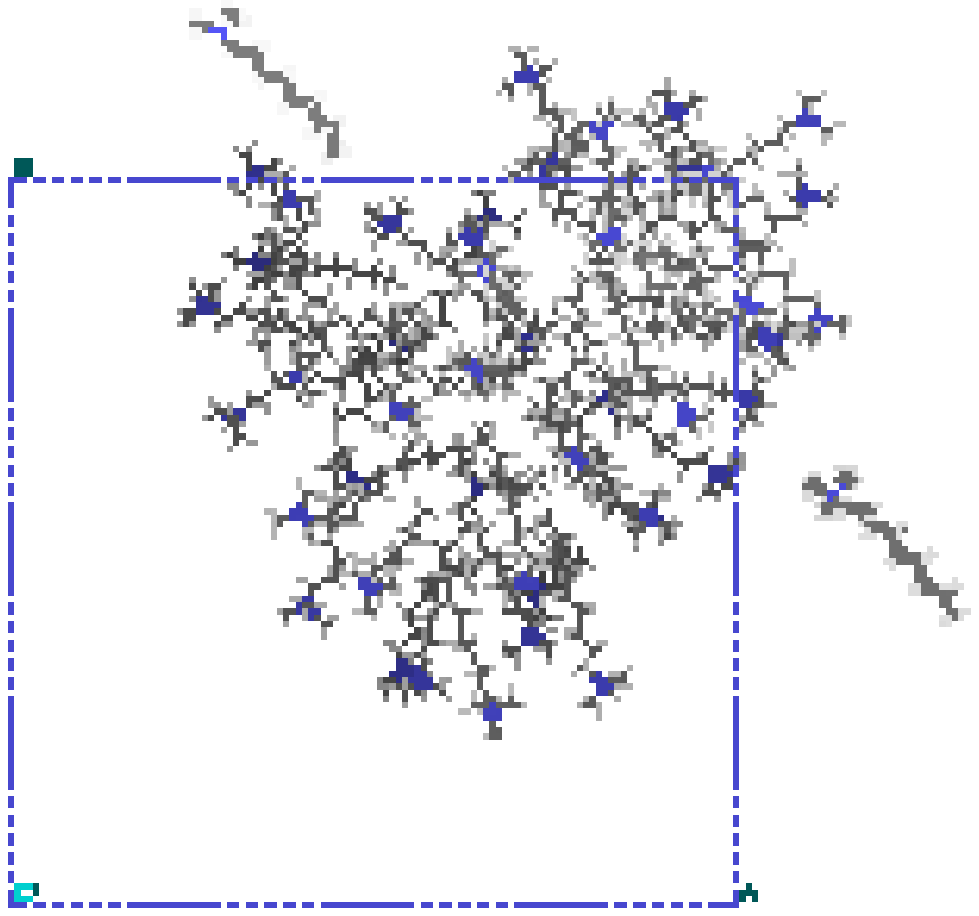}\\[2cm]
\includegraphics[scale=0.5]{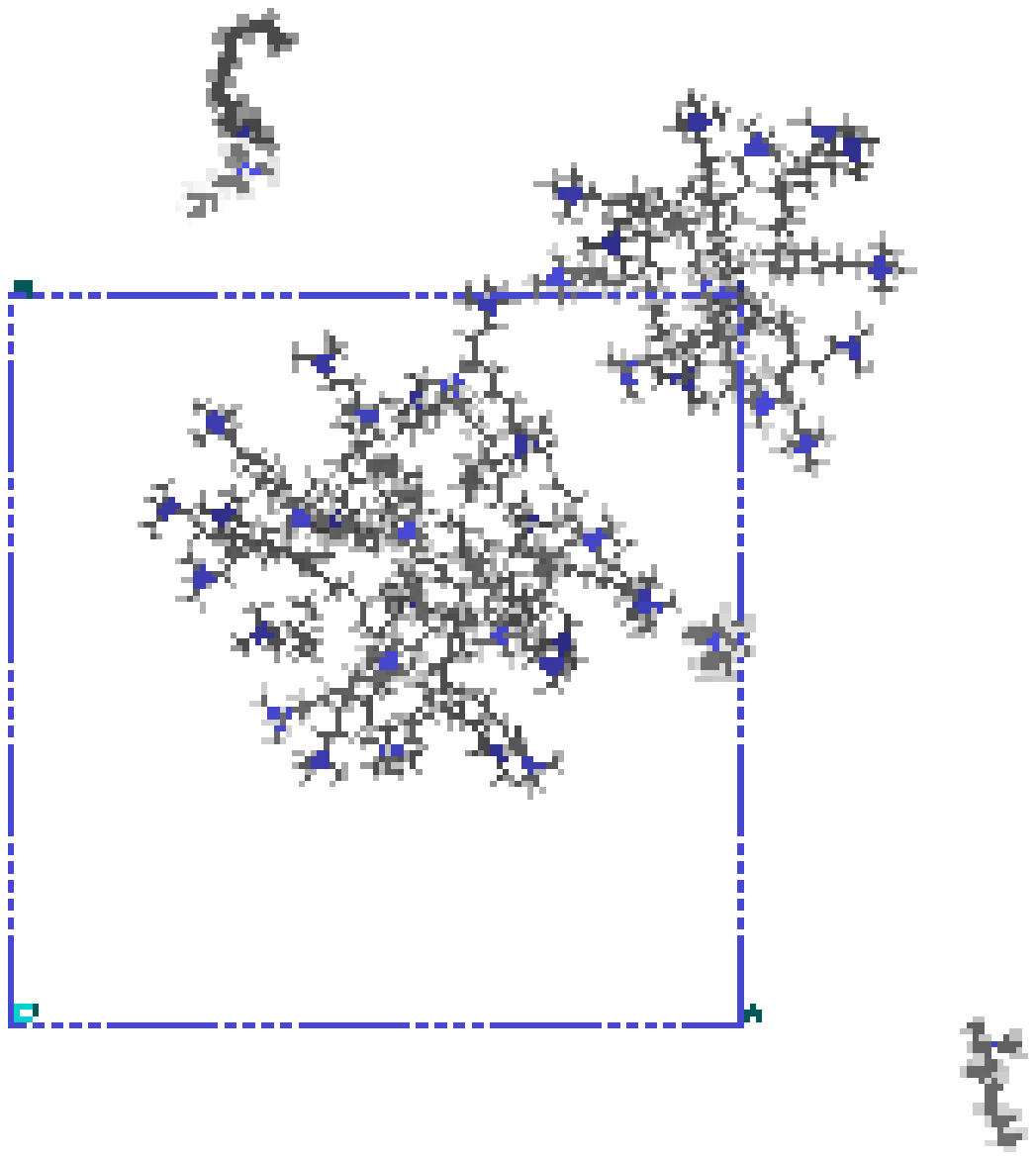}&
\includegraphics[scale=0.5]{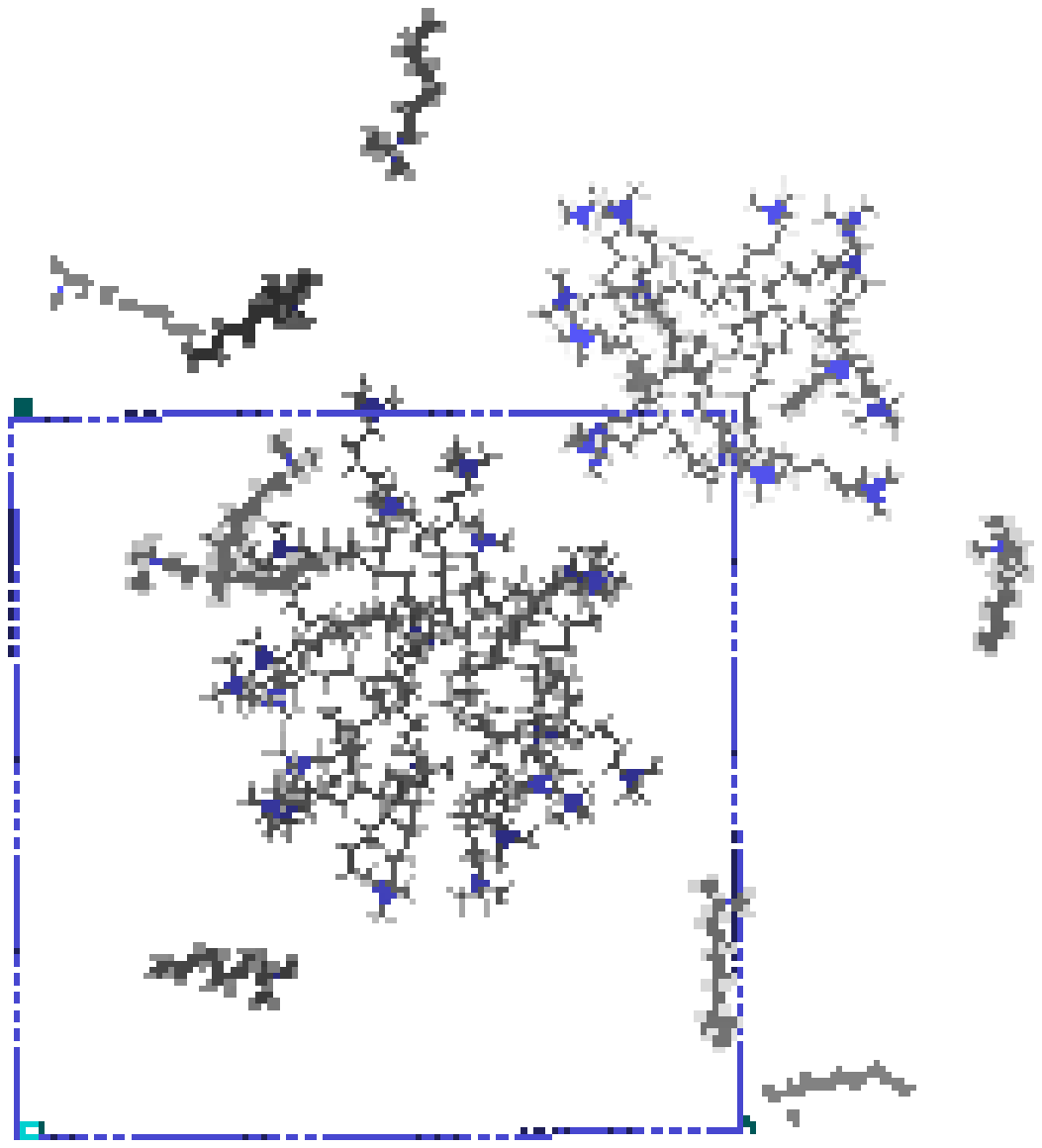}
\end{tabular}
\end{center}
\caption{\textbf{Instantaneous configurations of an initially
spherical $\rm \mathbf C_{9}TAC$ micelle after 50 ps (upper left), 600 ps
(upper right), 1100 ps (lower left) and 3100 ps (lower right) of
molecular dynamics. For clarity, water molecules and chloride counterions are not
displayed. In this and subsequent figures, the blue dotted lines
represent the edges of the periodic simulation box.}}
\label{micelle1}
\end{figure}

A spherical micelle was built with 48 $\rm C_{9}TAC$ surfactant
molecules surrounded by 2997 water molecules in a cubic box with sides
of length
70 $\rm \AA$. It took around $50$ ps of molecular dynamics at T=300 K
for the volume of the simulation cell to reach a stable
value $V=1.1$ $10^{-25}m^{3}$, corresponding to a density of $0.97$
$g/cm^{3}$. The surfactant concentration in the simulated
solution is thus $C=0.72$ $mol/l$. The values of the critical micelle
concentration (cmc) for similar
surfactant molecules like $n$-decyltrimethylammonium chloride
($\rm C_{10}TAC$) or $n$-dodecyltrimethylammonium chloride ($\rm
C_{12}TAC$) in water are
reported as $0.05$ $mol/l$~\cite{bacaloglu} and $0.061-0.065$
$mol/l$~\cite{mukerjee} respectively. Our simulations
are performed at a much higher concentration than the cmc; micelles
would thus be expected to form.\\
Figure~\ref{micelle1} shows snapshots
of the system at different times during the simulation. Water molecules
are not displayed in order to have a more detailed view of the
structure of the micelle. The calculated
values of the radius of gyration, the ratios of the lengths of the principal
axes, and the radius of the micelle at different timesteps,
are reported table~\ref{radiusTAC}. After $50$ ps, both the volume and
the total
energy of the system have reached stable values.
Figure~\ref{micelle1} gives evidence of the spherical shape of the
micelle. The polar
head groups are located on the micellar surface and are in direct
contact with water molecules; the alkyl chains are directed into the
hydrophobic core. Analogous views of the system at $600$ ps, $1.1$ ns and $3$
ns are displayed in figure~\ref{micelle1}. After $600$
ps, we observe that 3
surfactant molecules have left the micelle. At this stage, the micelle
contains $45$ molecules. Its shape is ellipsoidal rather than
spherical, as can be deduced from the ratios of the lengths of the
principal axes (see table~\ref{radiusTAC}).  After $1.1$ ns of simulation, the
micelle has broken down into two smaller spherical micelles, one containing
$29$ surfactant molecules and the other one $15$
molecules. The four remaining surfactant molecules are solubilized as
monomers in the water.
The observed mechanism through
which the micelle breaks into two smaller entities is as follows: the
initial micelle undergoes a structural change from spherical to
ellipsoidal, one of the principal axis becoming twice as long as the
other two. This ellipsoid looks like a small dumbbell (the density of
surfactant molecules at its middle is small).
Finally, the dumbbell separates into two spherical entities. A rapid
reorganisation of the surfactant molecules occurs after the
separation.\\
Experimental results of Imae \emph{et al.}~\cite{imae} on
several similar surfactants in aqueous solution indicate the presence
of spherical micelles with average aggregation numbers of $84$ for
$\rm C_{16}TAC$ ($n$-hexadecyl trimethylammonium chloride), $62$ for
$\rm C_{14}TAC$ ($n$-tetradecyl trimethylammonium chloride), and $44$ for
$\rm C_{12}TAC$ ($n$-dodecyl trimethylammonium chloride). An extrapolation
of their results would predict an average aggregation number of
approximately $25$ for $\rm C_{9}TAC$. At the end of the simulation
(figure~\ref{micelle1}), several surfactant molecules have left the
larger micelle while the smaller one has increased its size by
adsorption of one further monomer.
The two micelles finally contain $24$ and $16$ surfactant molecules,
with $8$ isolated surfactant monomers remaining in the solution.
Three nanoseconds of molecular dynamics simulation are clearly 
not long enough
to guarantee that
the system has reached thermodynamic equilibrium. At equilibrium, spherical micelles
are known to exhibit a size distribution rather than one particular
aggregation number. The size of a micelle thus evolves in
time, over a range given by the
polydispersity of the size distribution function. In this simulation,
one micelle has adsorbed one surfactant monomer while the other
micelle has desorbed a few monomers. 

\subsection{Infinite cylindrical micelle}
\begin{figure}[p]
\begin{center}
\begin{tabular}{cc}
\scalebox{0.5}{\includegraphics{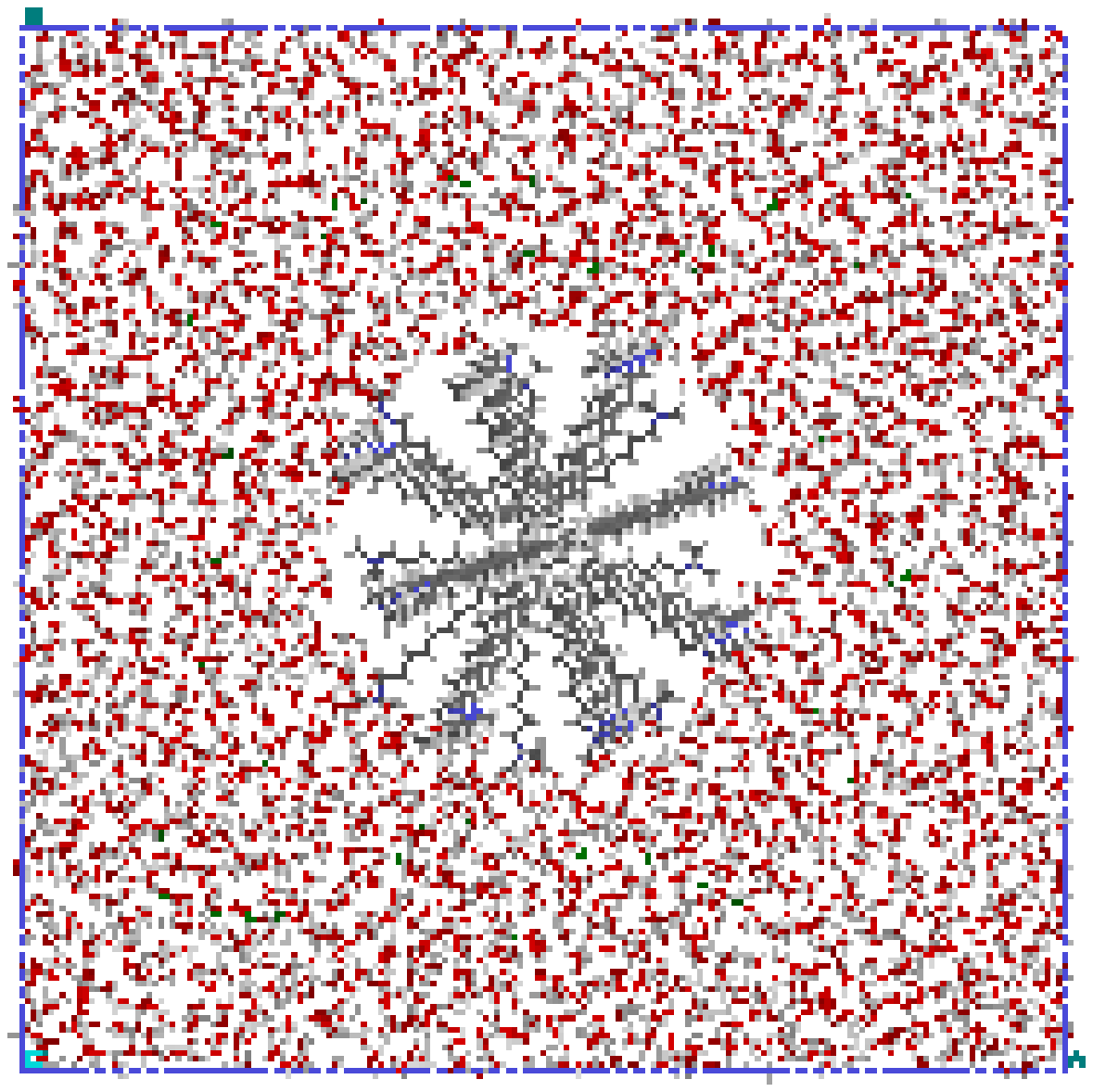}}&
\scalebox{0.5}{\includegraphics{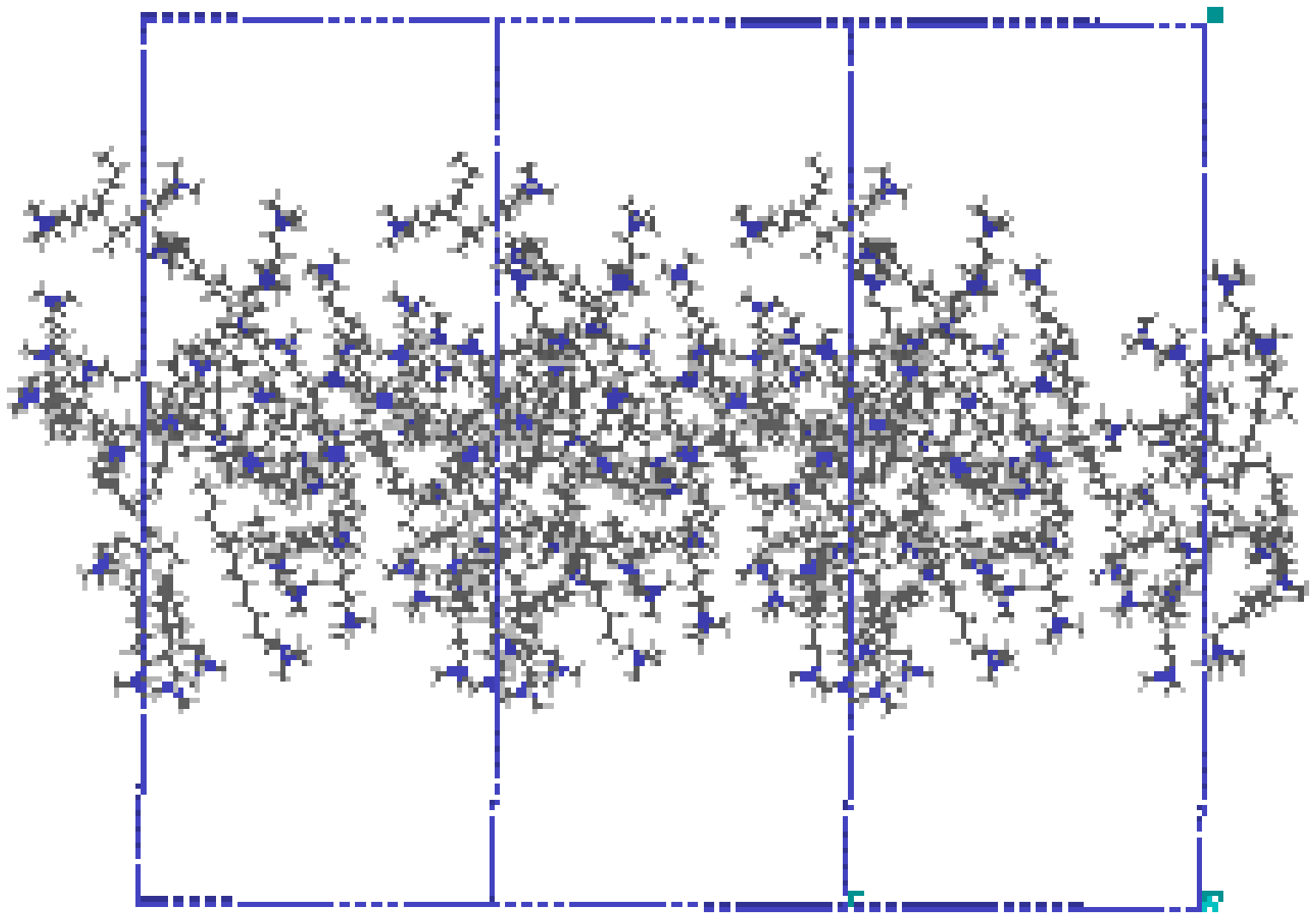}}\\
\scalebox{0.5}{\includegraphics{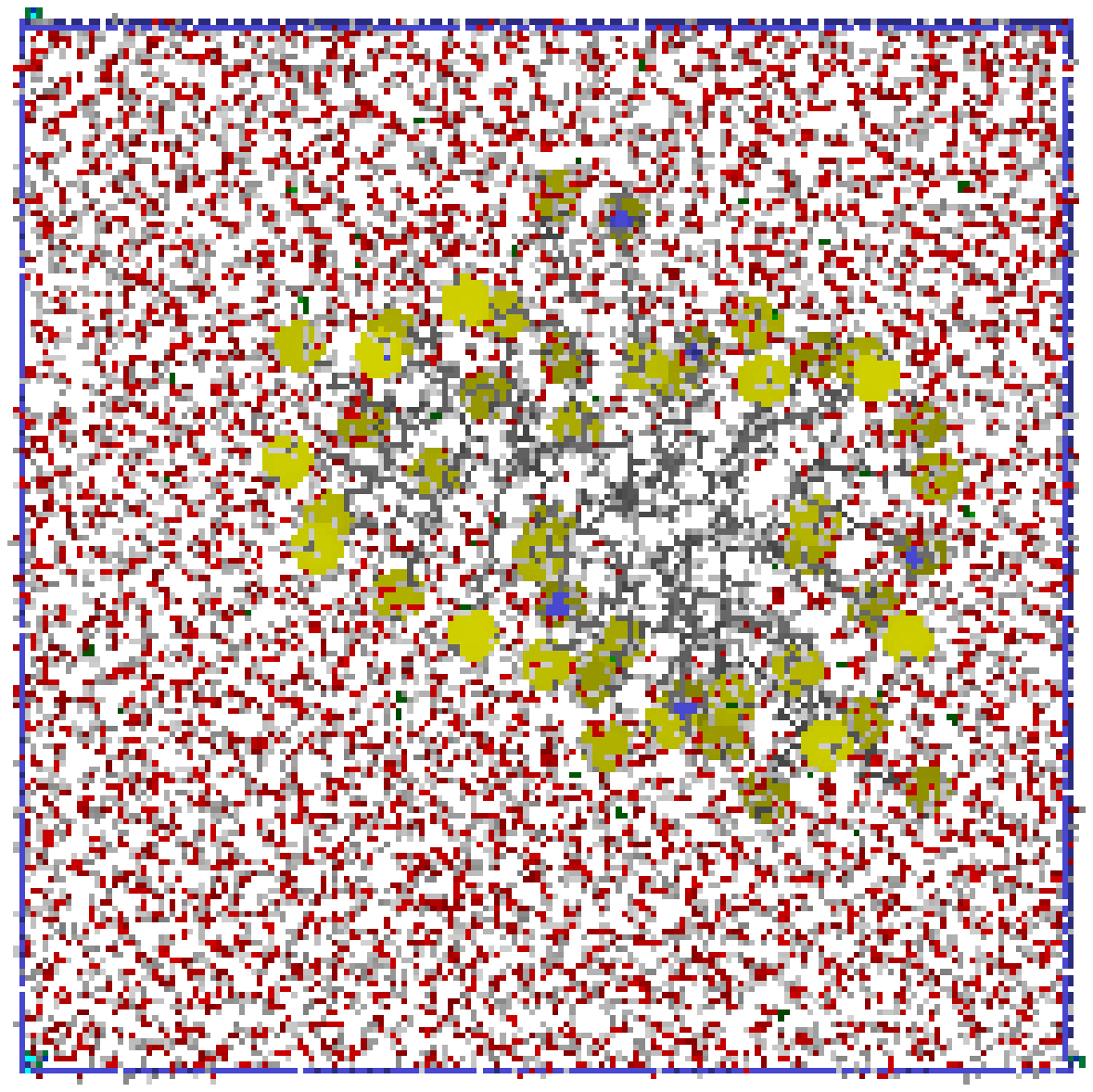}}&
\scalebox{0.5}{\includegraphics{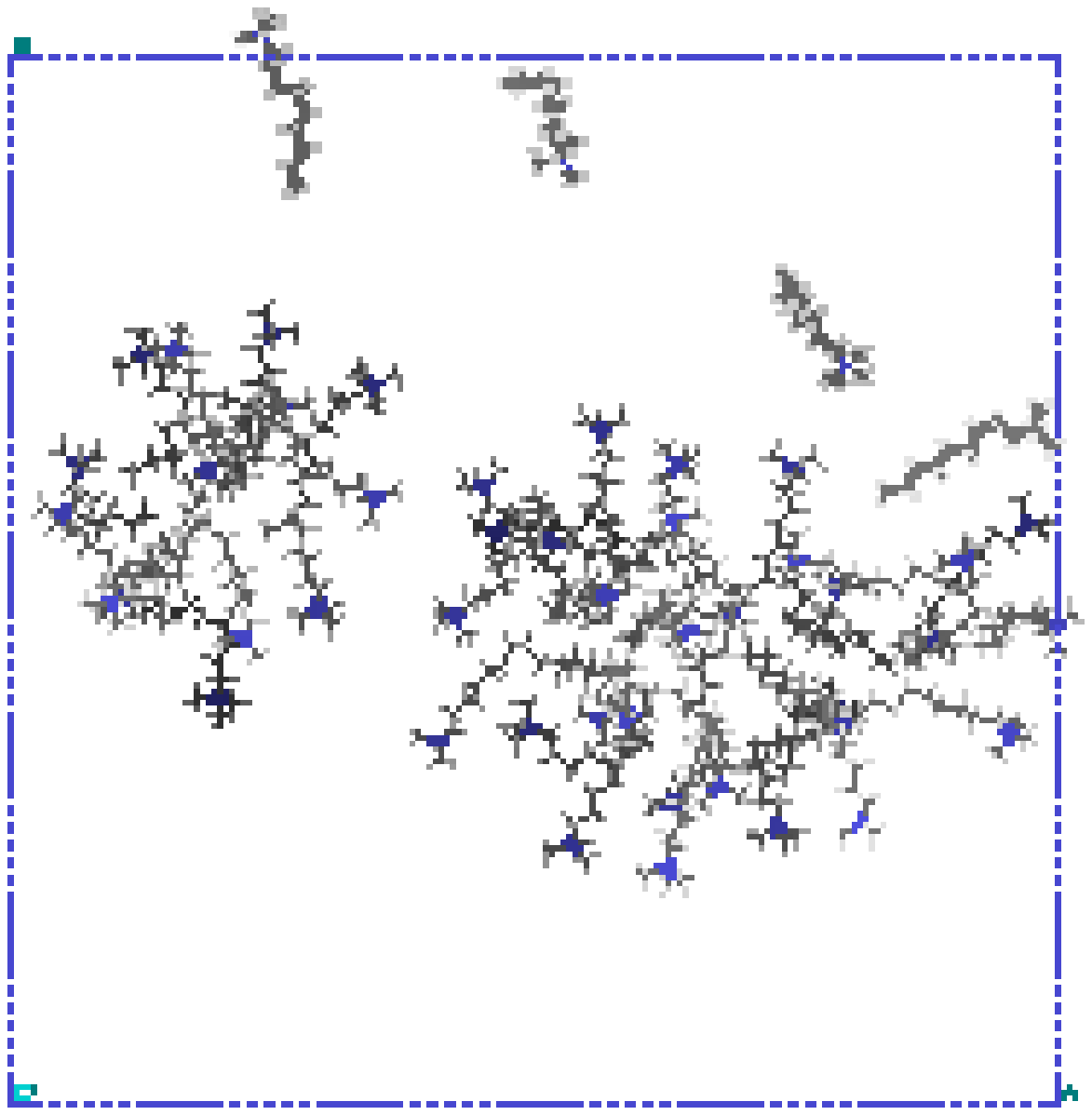}}
\end{tabular}
\end{center}
\caption{\textbf{Instantaneous configurations of an initially
infinite $\rm \mathbf C_{9}TAC$ cylindrical micelle at the beginning of the
simulation (upper left), after 200 ps
(upper right), 700 ps (lower left, also displaying in yellow the
Connolly surfaces of the cationic heagroups) and 3100 ps (lower right) of
molecular dynamics. In some snapshots, water molecules and chloride
counterions are not displayed.}}
\label{micelle2}
\end{figure}

A cylindrical micelle containing $48$ $\rm C_{9}TAC$ surfactant molecules was
constructed by putting together $6$ discs of $8$ molecules each. This
cylindrical micelle was surrounded by $4305$ water molecules. The
length of the initial cylinder was 30 $\rm \AA$ and 3D periodic boundary
conditions were applied in order to simulate infinite cylinders.
Figure~\ref{micelle2} shows the initial
configuration (viewed perpendicular to the axis of the cylinder), and
instantaneous configurations taken after $200$ ps, $700$ ps
and at the end of the simulation ($1.1$ ns); it takes
approximately $80$ ps to reach stable values of the volume of the
simulation cell and
the total energy. \\
The micelle evolves in time and some distortions
from the initial configuration appear quickly. After $200$ ps
of simulation, the micelle is still cylindrical, but the density of
molecules along the axis of symmetry of the cylinder no longer appears to
be uniform: the surfactant cations clump together,
forming high density regions while other regions of the worm exhibit
a small number of surfactant molecules. The latter correspond to
``weaker'' points, i.e. preferential zones for fragmentation of the
worm, with weaker hydrophobic tail interactions. Such a phenomenon signals the incipient
break-up of the infinite cylinder, as is
clearly seen in figure~\ref{micelle2} after $700$ ps of simulation:
the initial cylinder has expelled a small spherical micelle comprising $15$
surfactant molecules. The characteristic value of this spherical
micelle's radius of
gyration and the lengths of its principal
axes are reported in table~\ref{radiusTAC}, providing evidence of its
spherical shape. The rest of the cylindrical micelle  contains $30$
cations and exhibits a non-spherical structure. At the end of the
simulation ($t=1.1$ ns), the small spherical micelle contains $14$
surfactant cations, and the other one $30$ monomers. The
remaining four monomers are located in the aqueous solvent. The
different characteristic shapes of these micelles are reported
table~\ref{radiusTAC}. This state is similar to the state reached in
the previous simulation after fragmentation of the spherical micelle into
two micelles of sizes 15 and 30. We thus conclude that the time evolution of
this state would probably produce the same features as those described previously,
tending to thermodynamic equilibrium.

\subsection{Infinite cylindrical micelle with electrolyte}
\begin{figure}[p]
\begin{center}
\begin{tabular}{cc}
\scalebox{0.5}{\includegraphics{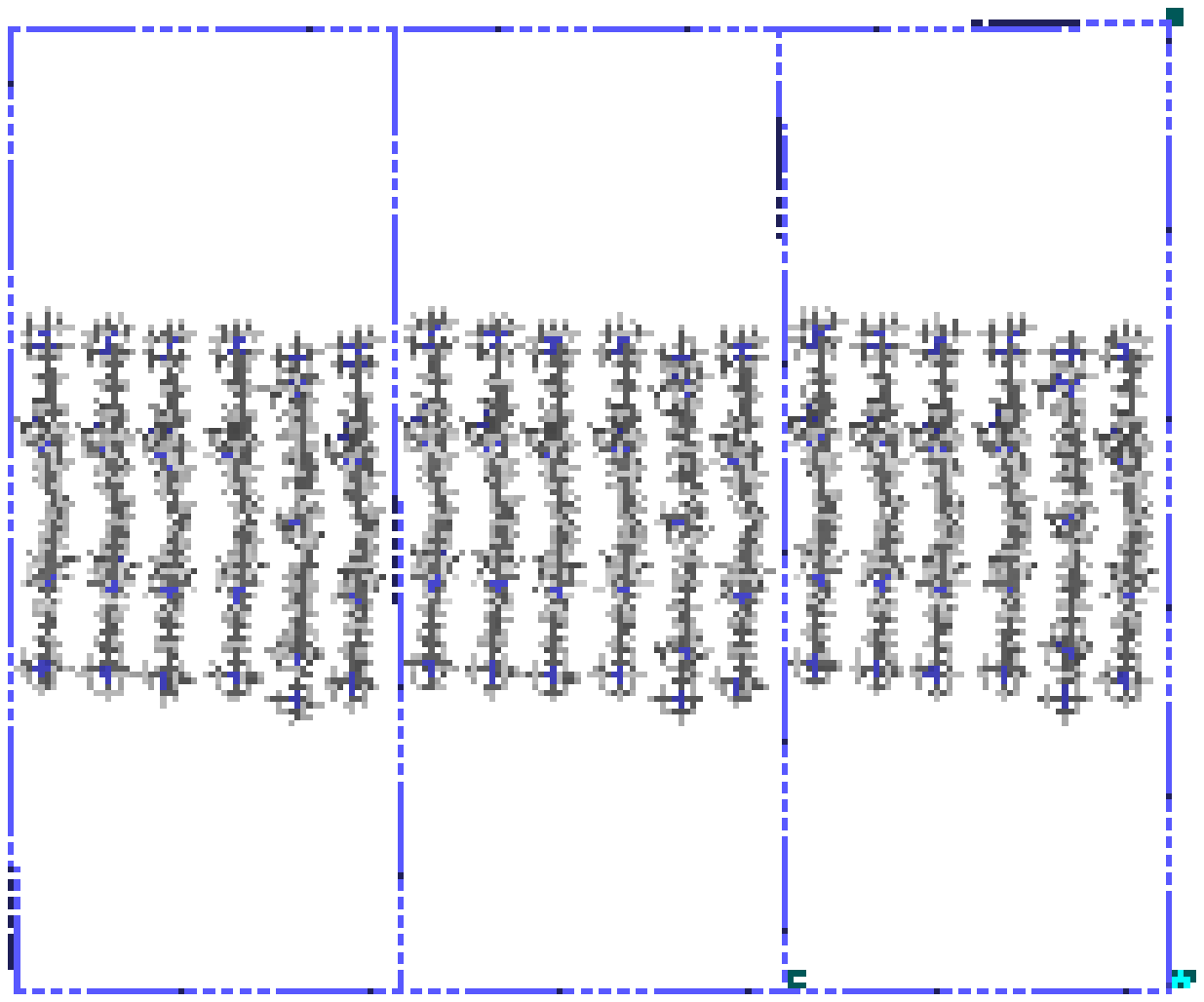}}&
\scalebox{0.5}{\includegraphics{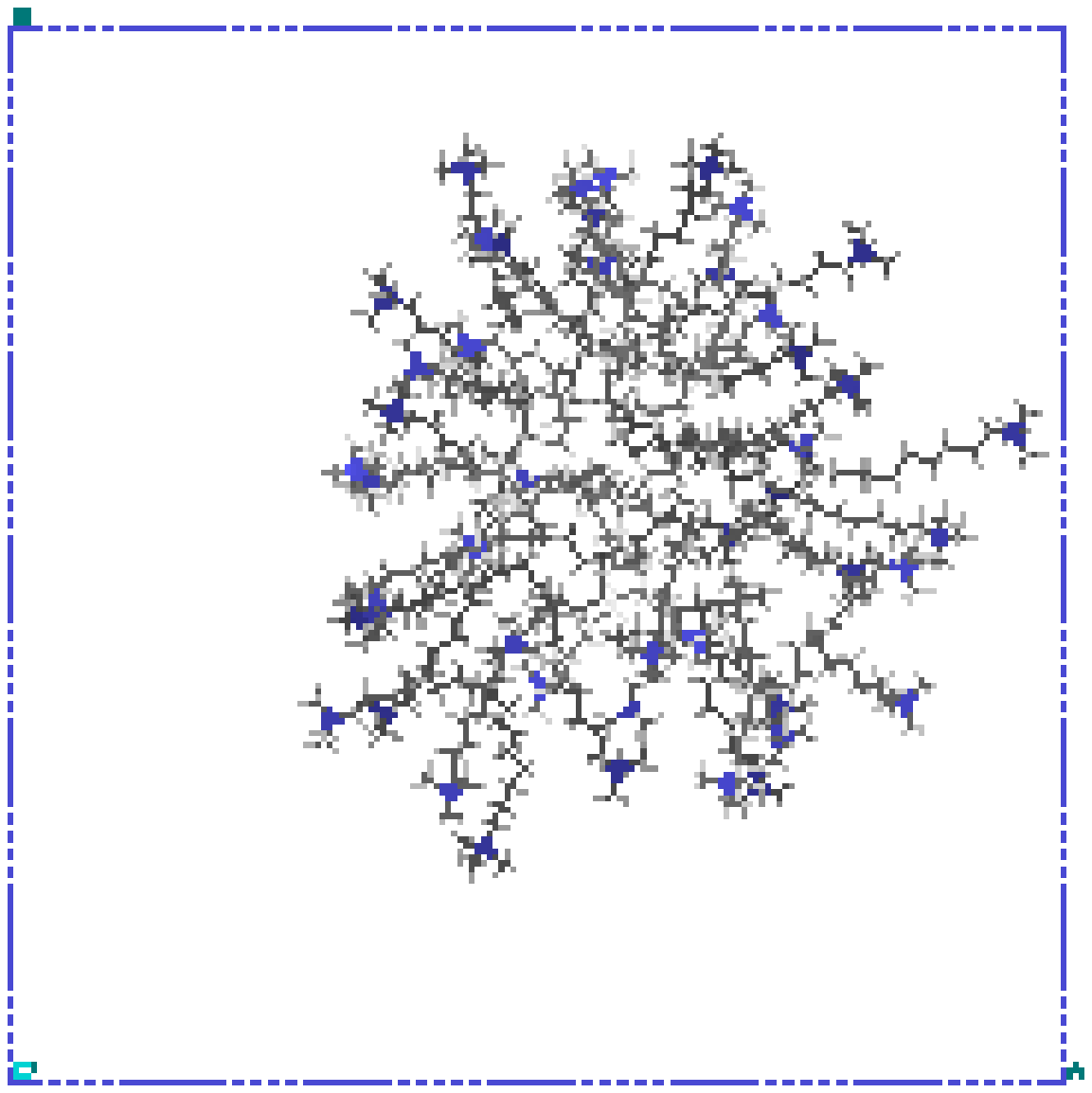}}\\
\scalebox{0.5}{\includegraphics{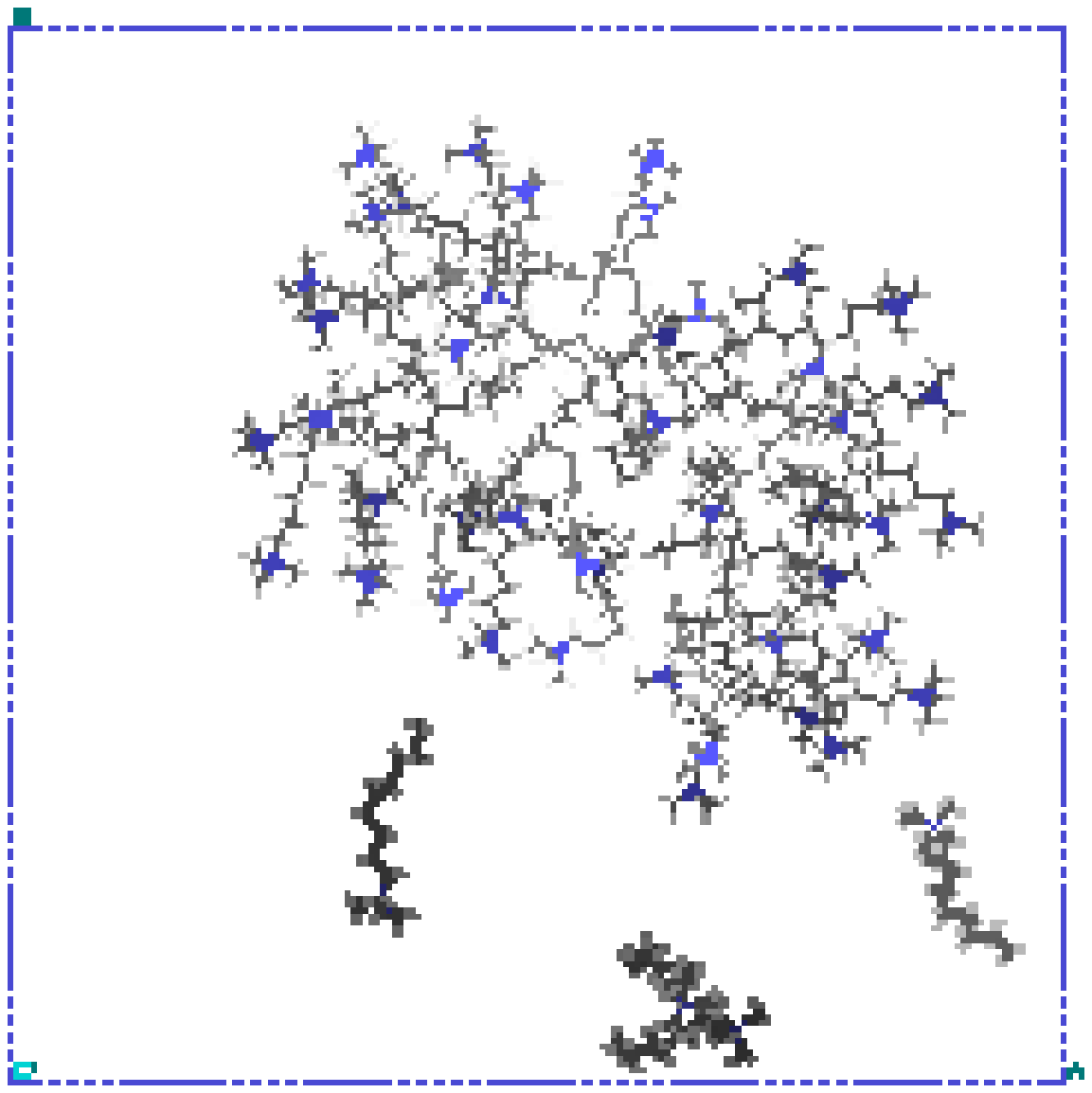}}&
\scalebox{0.5}{\includegraphics{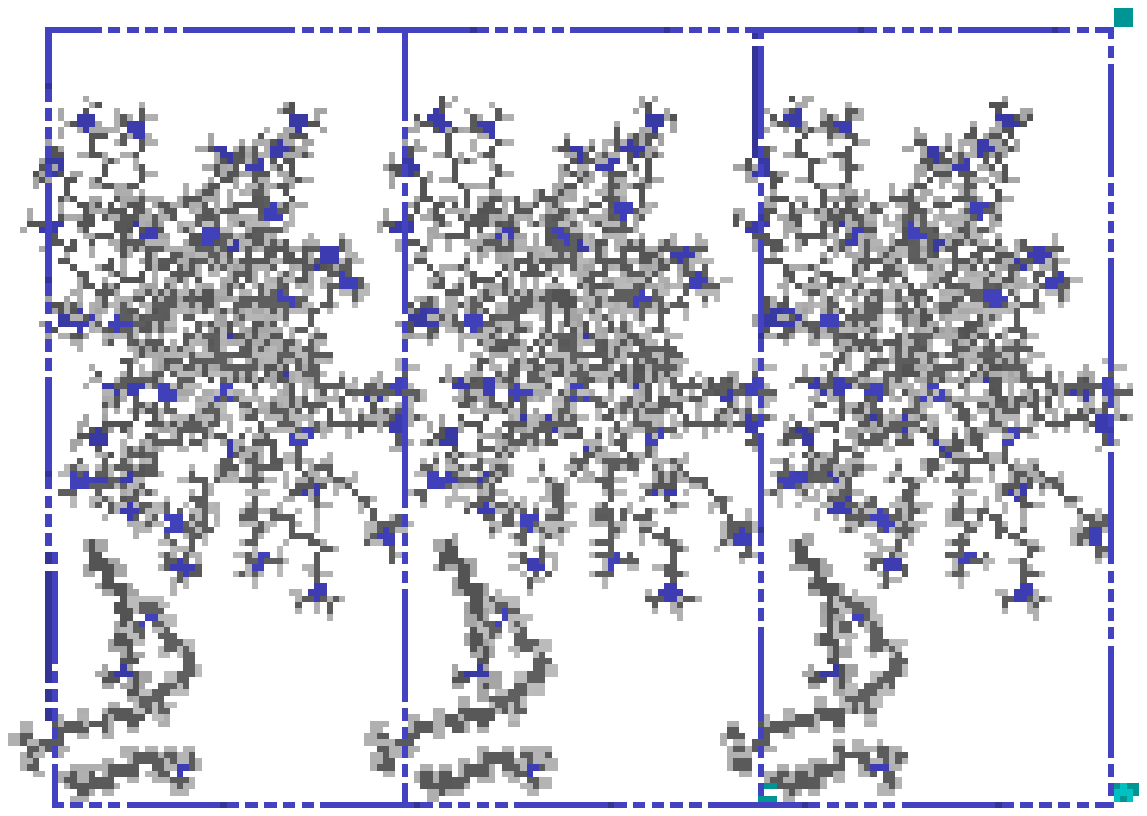}}
\end{tabular}
\end{center}
\caption{\textbf{Instantaneous configurations of an initially
infinite $\rm \mathbf C_{9}TAC$ cylindrical micelle with added electrolyte
at the  beginning of the molecular dynamics simulation (upper left),
after 100 ps (upper
right), and 2500 ps (lower part with view perpendicular (left) and
parallel (right)
to the axis of the small cylinder). Water molecules and chloride
counterions are not displayed.}}
\label{micelle3}
\end{figure}

The addition of electrolyte to an aqueous solution of surfactant
micelles is known empirically to preferentially
stabilize the cylindrical shape~\cite{israelachvili,israel2}. For a cationic
surfactant, the negative ions of the added salt
associate with the positively charged head groups of the surfactant.
Such associations reduce the strong electrostatic repulsion between
neighbouring head groups that exists in the cylindrical structure where
the latter lie closer together than they do on the surface of a
sphere (thus the packing parameter $P$ increases due to a decrease in
the surface area per head group $a$). Our aim here was to investigate whether $\rm C_{9}TAC$
cylindrical micelles
could indeed be stabilized by adding an electrolyte.\\
The model was constructed in the same way as the previous one
for an infinite cylindrical micelle but now $\rm Na^{+},Cl^{-}$ ion-pairs were
added to the solution. Specifically, the system was composed of $48$ surfactant
cations (and $48$ chloride counterions), $2997$ water molecules, and
$100$ sodium chloride ion pairs, each ion being placed at random
within the water molecules of the solvent. The concentration of
chloride anions is thus $C=1.64$ $mol/l$. A view perpendicular to the
principal axis of the cylinder is
shown in figure~\ref{micelle3}.
After $100$ ps of MD simulation (figure~\ref{micelle3}), the infinite
micelle has broken down into a
finite micelle. Its shape remains roughly cylindrical, as evidenced by
the ratios of the principal axes reported in table~\ref{radiusTAC}. In
order to compare the
stability of this structure with that of the spherical
micelles produced in the absence of electrolyte, the simulation was
run up to
$2.5$ ns. At the end of the simulation (figure~\ref{micelle3}), the
micelle still exhibited a cylindrical shape, containing $44$ surfactant
molecules (four monomers left the cylinder and were individually dissolved
in the water): it is an example of a
finite, rod-like micelle (self-assembled structures which have been
observed experimentally). However, we cannot confirm that this system
is near equilibrium, owing to the lack of either longer-time
simulation data or relevant experimental results.

\subsection{Random monomer distribution}
\begin{figure}[p]
\begin{center}
\begin{tabular}{cc}
\scalebox{0.5}{\includegraphics{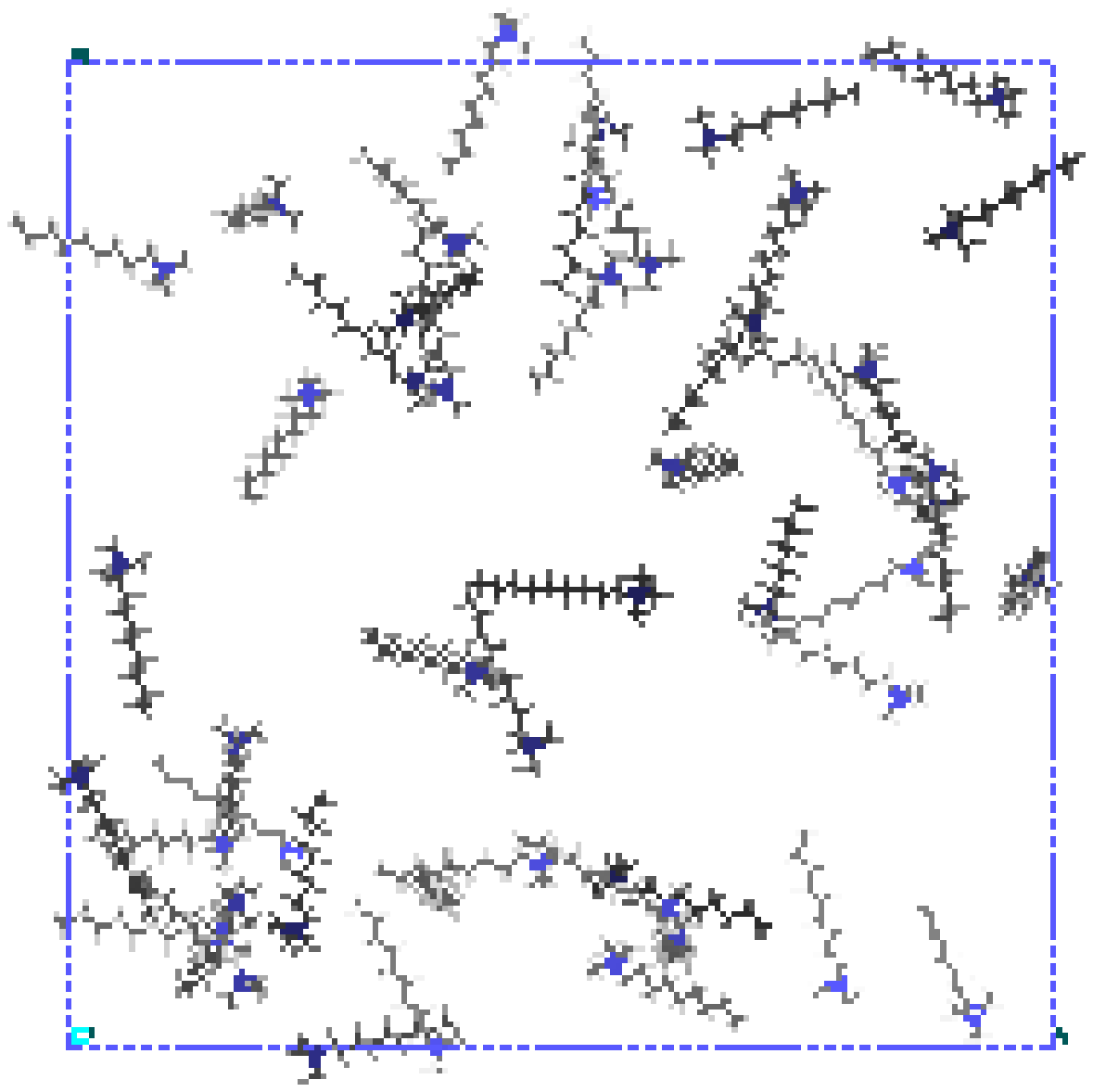}}&
\scalebox{0.5}{\includegraphics{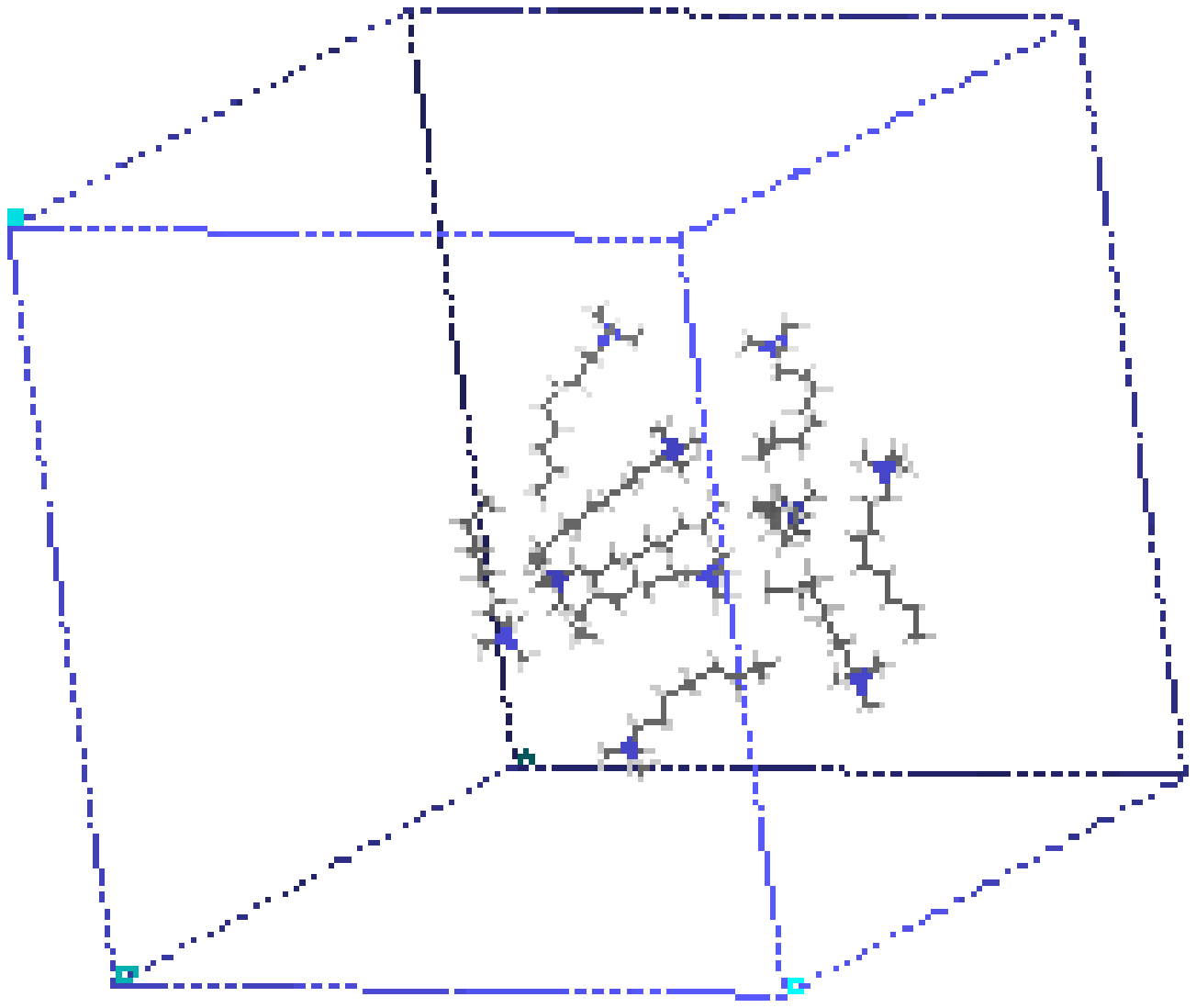}}\\[1cm]
\scalebox{0.5}{\includegraphics{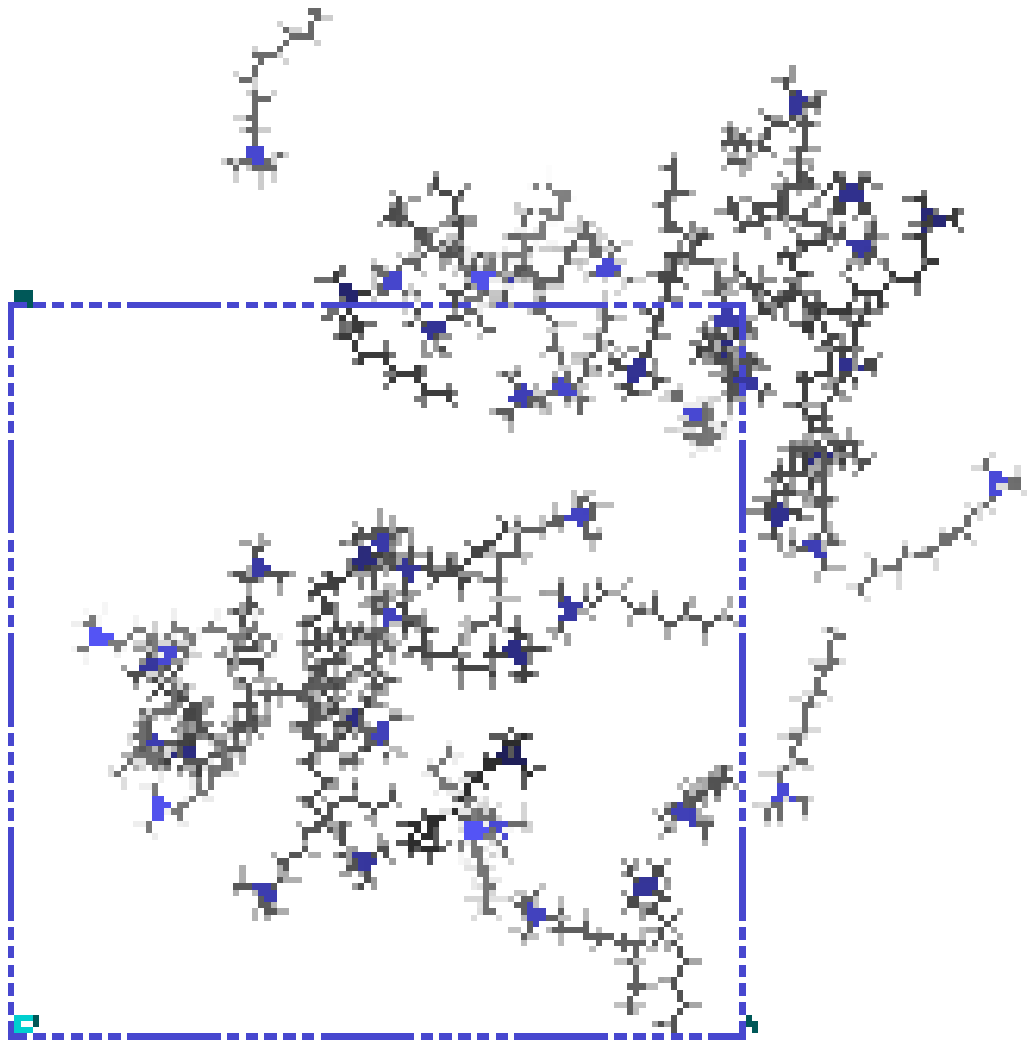}}&
\scalebox{0.5}{\includegraphics{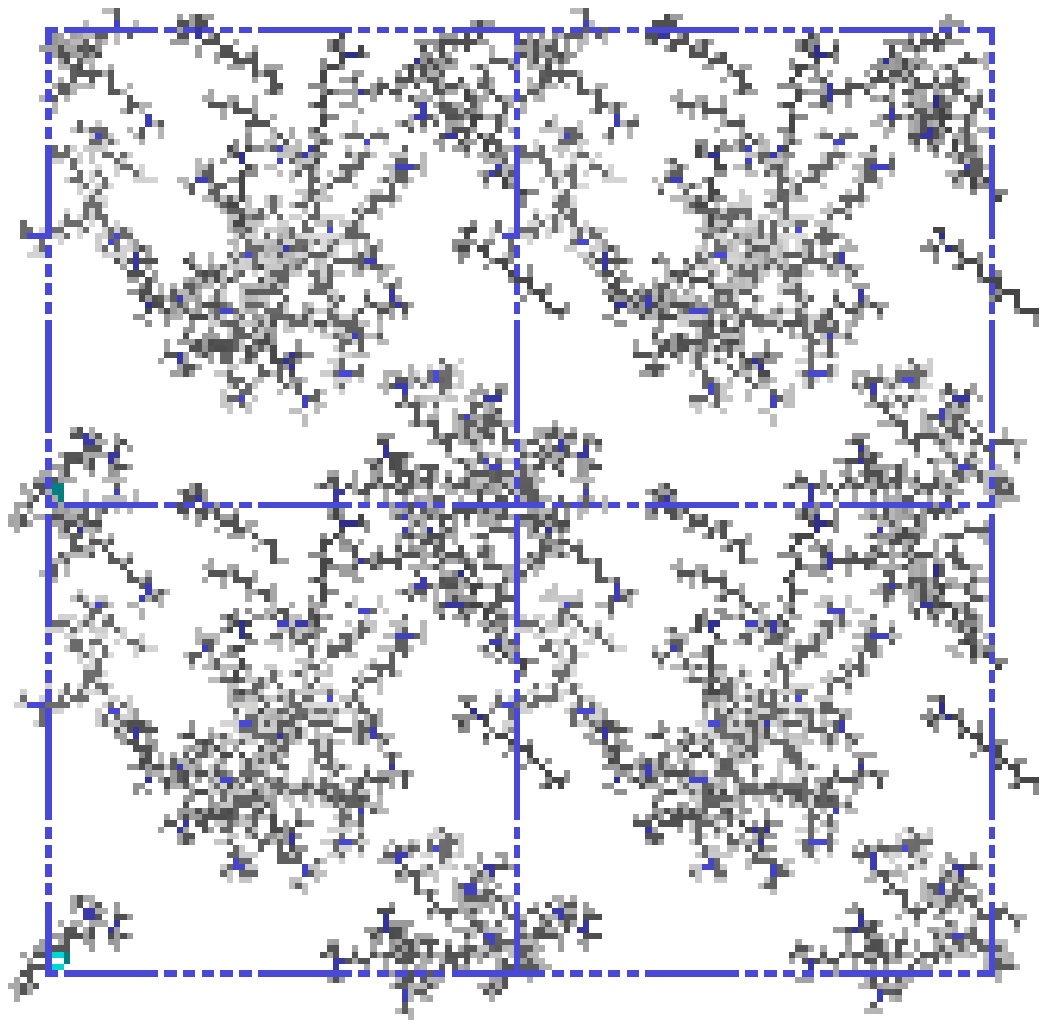}}
\end{tabular}
\end{center}
\caption{\textbf{Instantaneous configurations of an initially random
distribution of $\rm \mathbf C_{9}TAC$ monomers at a concentration
above the expected critical micelle concentration at the beginning of the
simulation (upper left), after 75 ps (upper right), at 200 ps (lower
left) and at 900 ps (lower right) of molecular dynamics. Only the
surfactant molecules of interest are displayed. After 200 ps
of molecular dynamics, two spherical micelles are already formed.}} 
\label{micelle4}
\end{figure}

A starting configuration of $48$ surfactant and $2997$ water molecules
was prepared, the surfactant cations and chloride counterions being
randomly distributed
throughout the system (see figure~\ref{micelle4}). The simulation
cell was a cubic box with sides of length $70$ $\rm \AA$, corresponding
to an initial surfactant density of $0.3$ $\rm g/cm^{3}$. The simulation was run up to
$900$ ps in order to obtain information on the mechanism of surfactant
micelle formation. At $t=75$ ps, the system has already become
inhomogeneous with some low and high density regions
(figure~\ref{micelle4}). In the high density regions, no particular
arrangement of the surfactant molecules is discernible. A structural
aggregation between the surfactant particles is seen at $t=200$ ps
(figure~\ref{micelle4}), leading to the appearance of two micelles.
Each micelle is composed of approximately $15$ molecules. By examining
several instantaneous configurations between $t=75$ and $t=200$
ps, we can obtain insight into the dynamics of 
spherical micelle formation at
the molecular level. It appears that in the first stage (between 0 and
100 ps), the surfactant molecules approach one another, forming
aggregates without any well-defined organisation, both from a
translational and rotational point of view
(figure~\ref{micelle4}). The size of
these disordered micelles is small ($\sim 10$ molecules). In the
second step these random aggregates rearrange to form spherical
micelles.
The driving force for this rearrangement appears to be the
minimisation of the repulsive
interactions between head groups together with the hydrophobic
attractions between
the hydrocarbon tails. During this structural rearrangement, the
aggregation number of the micellar clusters increases via the
addition of small clusters of surfactant (typically 2 or 3 molecules).\\
Finally, by the end of the simulation, both micelles exhibit a spherical
shape, containing $15$ and $17$ surfactant molecules respectively.
Their characteristic radii are reported in table~\ref{radiusTAC}. A
small number of monomers remain solvated in the surrounding aqueous
medium. The sizes of the micelles are consistent with the assumption
of a mean aggregation number of around 15-20.

\section{Structure and dynamics of EMAC micelles}
\subsection{Spherical micelle}
\begin{figure}[p]
\begin{minipage}[t]{14.0cm}
\begin{minipage}[t]{7.0cm}
\centering
\scalebox{0.5}{\includegraphics{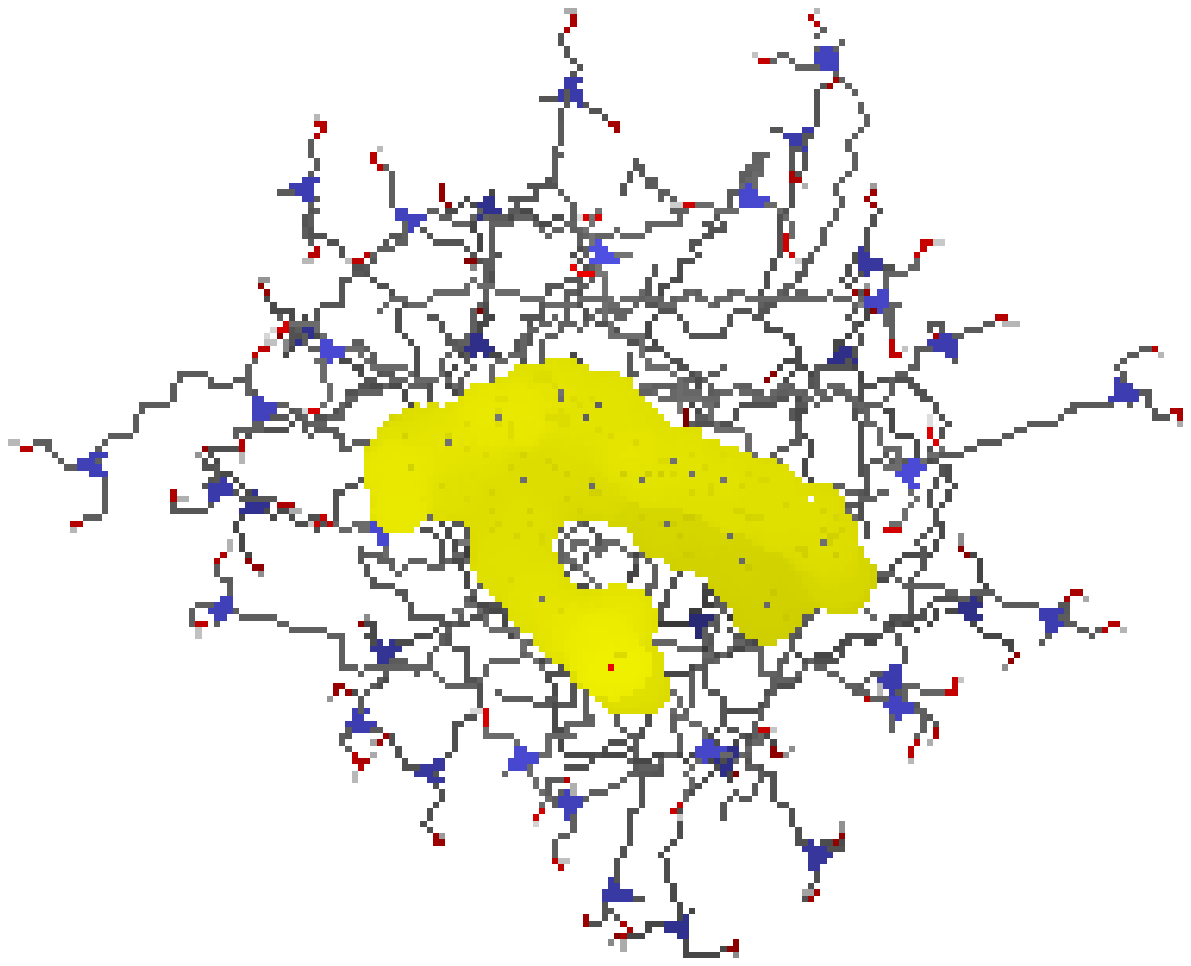}}
\end{minipage}
\begin{minipage}[t]{7.0cm}
\centering
\scalebox{0.5}{\includegraphics{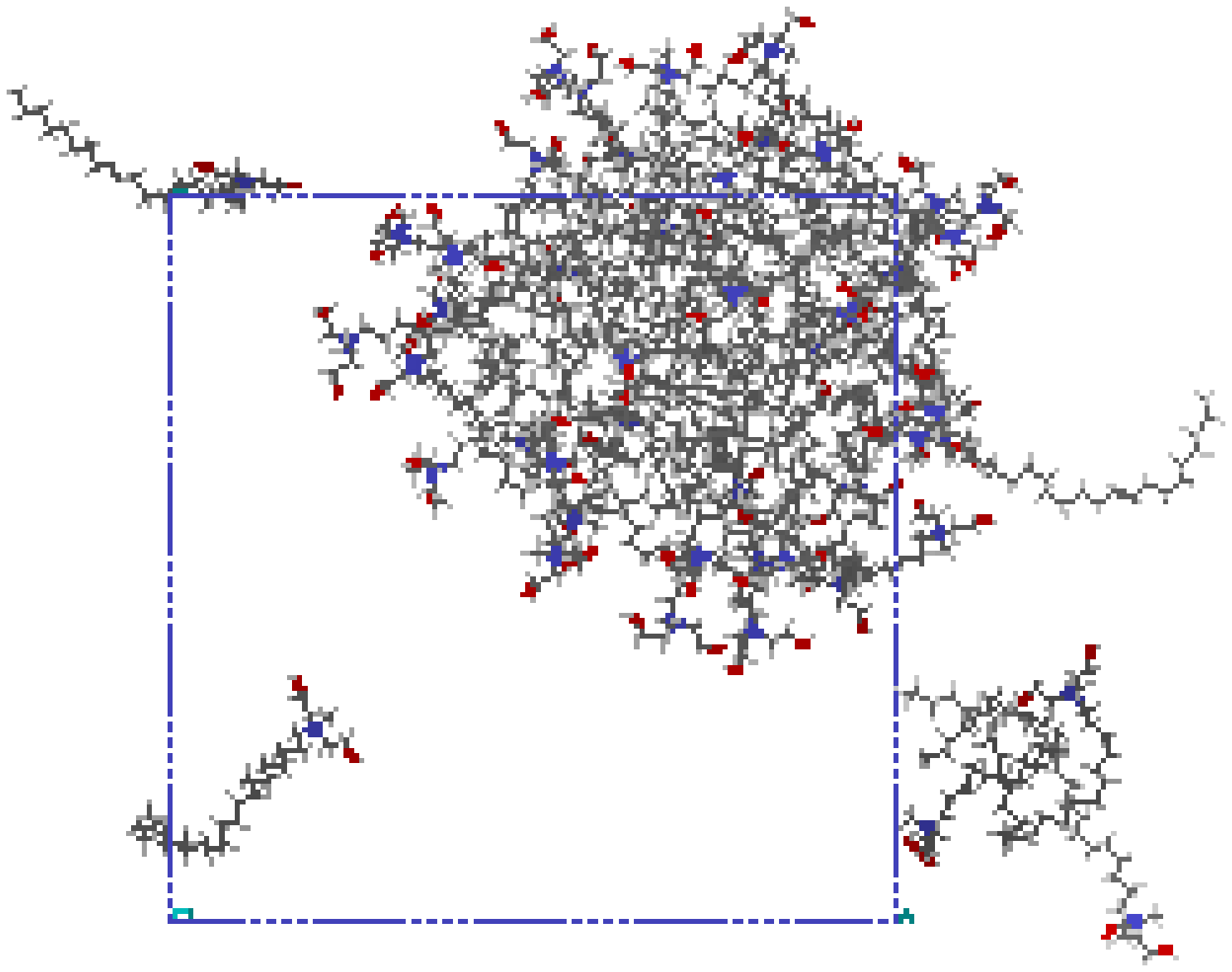}}
\end{minipage}
\end{minipage}
\caption{\textbf{Instantaneous configurations of an initially
spherical EMAC micelle after 500 ps (left), and at the end (right) of a
molecular dynamics simulation ($1.05$ ns). The first snapshot reveals
the surface
adsorption of an EMAC dimer, displayed in terms of its Connolly
surface (in yellow); the second image shows the sphericity of the
micelle, after the adsorption of the dimer; the dimer has remained on
the surface of the cluster (not displayed here).}}
\label{micelle5}
\end{figure}

A simulation cell was constructed, comprising $40$ EMAC molecules
within a spherical micelle, together with
$8$ solvated monomers, $48$ chloride counterions and 3497 water
molecules. Molecular dynamics was performed on this system up to
$1.05$ ns. It took $120$
picoseconds for the volume of the simulation cell and the total energy to reach
stable values (the surfactant concentration is equal to $0.54$
$mol/l$). At this point in the simulation, the characteristic
dimensions of the spherical micelle are as reported in table~\ref{radiusCF}. During
the entire simulation, the micelle maintained its spherical shape. Contrary to
the case of $\rm C_{9}TAC$ spherical micelles, no fragmentation occurred
within this timescale. This is possibly due to the fact that the
expected aggregation number for a spherical EMAC micelle may be
greater than $40$~\cite{hughes}.
Nevertheless, two surfactant molecules initially dissolved
in water approached the micelle and adhered to its surface. This
phenomenon is shown in
figure~\ref{micelle5}, where the two adsorbing molecules are
highlighted. We
can see that the two molecules remain close to each other like a
dimer, even after
their adsorption at the micellar surface. Their hydrophobic tails are not directed toward
the centre of the micelle; they rather remain in close proximity to
water along their entire length. Figure~\ref{micelle5} shows a view
of the final
configuration after $1.05$ ns. The shape of the micelle is spherical,
and 6 surfactant
molecules remain dispersed in water.

\subsection{Infinite cylindrical micelle}
\label{cylemac}
\begin{figure}[p]
\begin{minipage}[t]{14.0cm}
\begin{minipage}[t]{7.0cm}
\centering
\scalebox{0.5}{\includegraphics{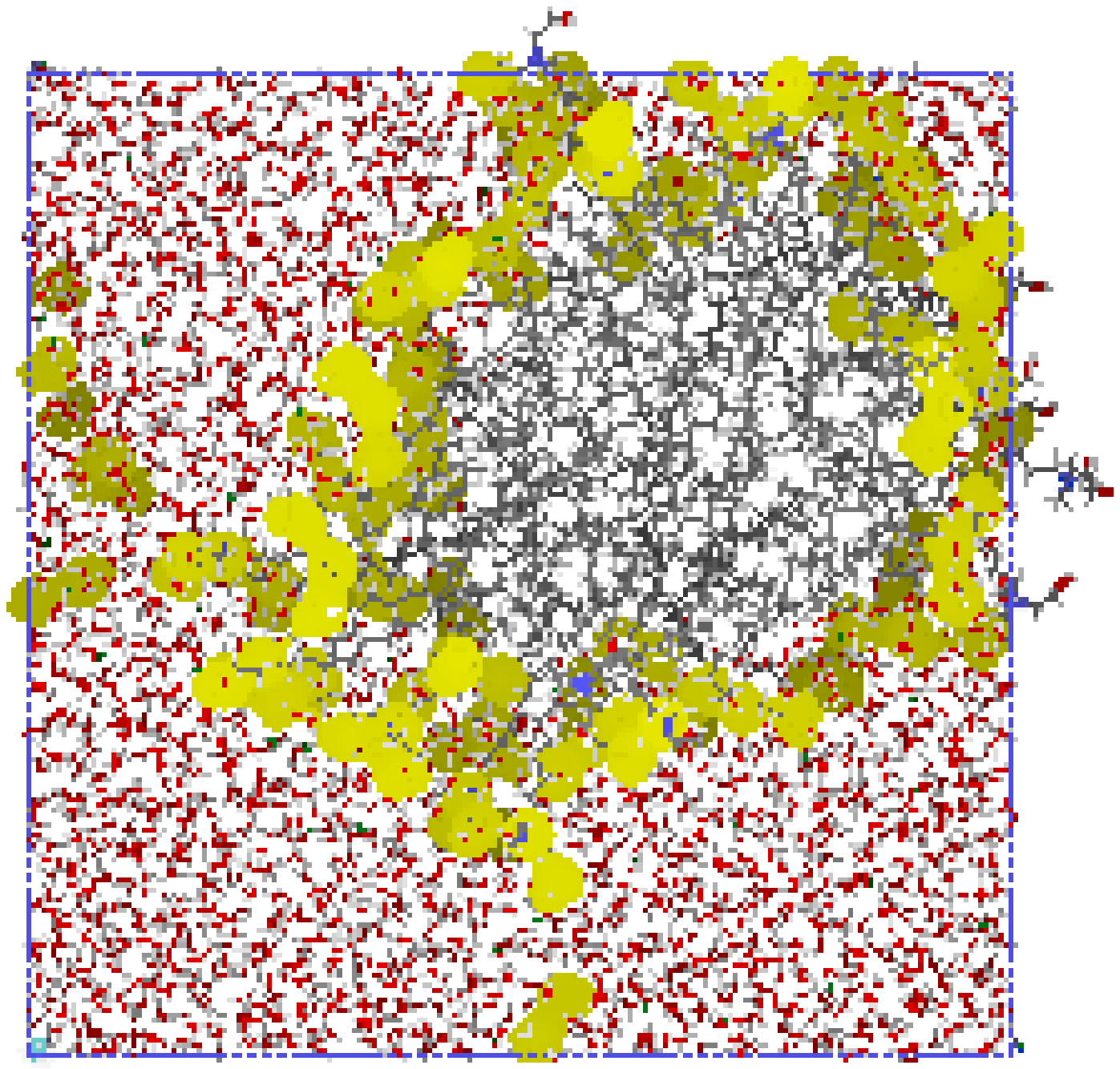}}
\end{minipage}
\begin{minipage}[t]{7.0cm}
\centering
\scalebox{0.5}{\includegraphics{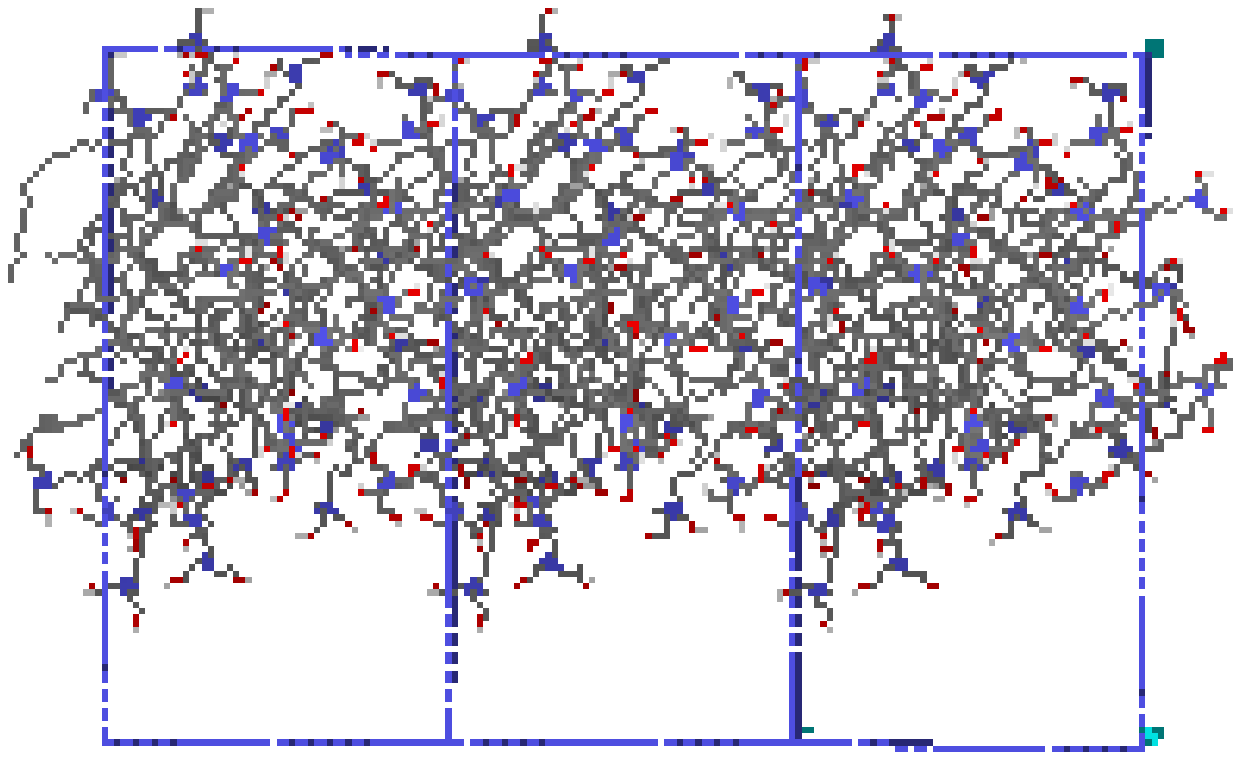}}
\end{minipage}
\end{minipage}
\caption{\textbf{Instantaneous configurations of an initially
infinite EMAC cylindrical micelle at the end of the molecular dynamics
simulation
(1850 ps) with views perpendicular (left) and parallel (right) to the
axis of the cylinder. In the left hand image, the cationic head groups
are displayed as Connolly surfaces (in yellow).}}
\label{micelle6}
\end{figure}

A cylindrical micelle of EMAC molecules was constructed by combining
five discs of 10 molecules each. Due to the periodic boundary
conditions, the cylinder is effectively infinite in the direction of
its principal axis.
This micelle was
surrounded by $3119$ water molecules and $50$ chloride ions. The
system took $100$ ps to equilibrate (that is, to attain stable values
of the simulation cell volume and
the total energy). Over a total period of $1.85$ ns of molecular
dynamics, the micelle retained
its cylindrical shape. No surfactant monomers were present in the surrounding
water at the beginning of the simulation and neither
desorption of monomers, nor any other form of fragmentation was observed:
the micelle remained an infinite
cylinder throughout. Figure~\ref{micelle6} shows a projection of the
system on
a plane perpendicular to the axis of the cylinder, at the end of the
simulation. Water molecules are also shown. The
cross section of the micelle is more ellipsoidal than circular, as
can be deduced from the values of the ratios of the lengths of the
principal axes (see table~\ref{radiusCF}). Moreover, no penetration of
water molecules in the micelle core was observed; water molecules remain at the micellar
surface, in close proximity to and surrounding the head groups. The core
of the micelle is indeed
completely anhydrous. The dimensions of a cross section of the micelle are
reported in table~\ref{radiusCF}.\\
A snapshot of the
system is also shown in figure~\ref{micelle6}, perpendicular to the axis of
the cylinder, where three images of
the periodic simulation cell are
displayed. The micelle exhibits regions of varying density along its
principal axis. Some local regions of the worm are narrower than
others, corresponding to ``weaker'' points. With our enhanced
understanding of the behaviour of $\rm C_{9}TAC$ micelles, this
could be interpreted as the incipient site of rupture of the cylinder into
smaller spherical micelles. Nevertheless, we believe that this system
is still at some distance from equilibrium, because the final state in this
simulation is very different from the one obtained in the previous simulation
(spherical micelle) while the compositions of both systems are more or less identical.

\subsection{Random EMAC monomer distribution}
\begin{figure}[p]
\begin{minipage}[t]{14.0cm}
\begin{minipage}[t]{7.0cm}
\centering
\scalebox{0.5}{\includegraphics{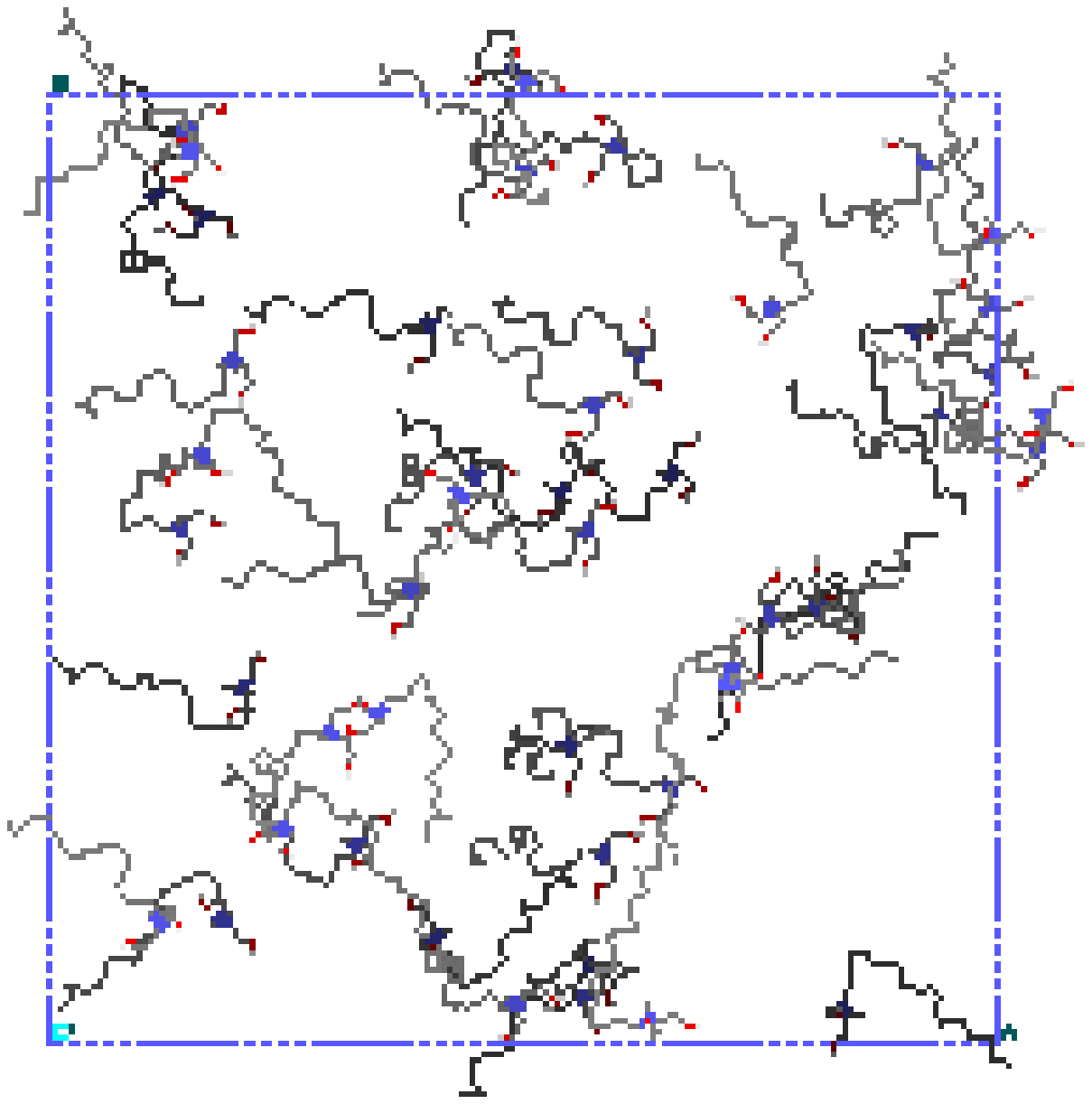}}
\scalebox{0.5}{\includegraphics{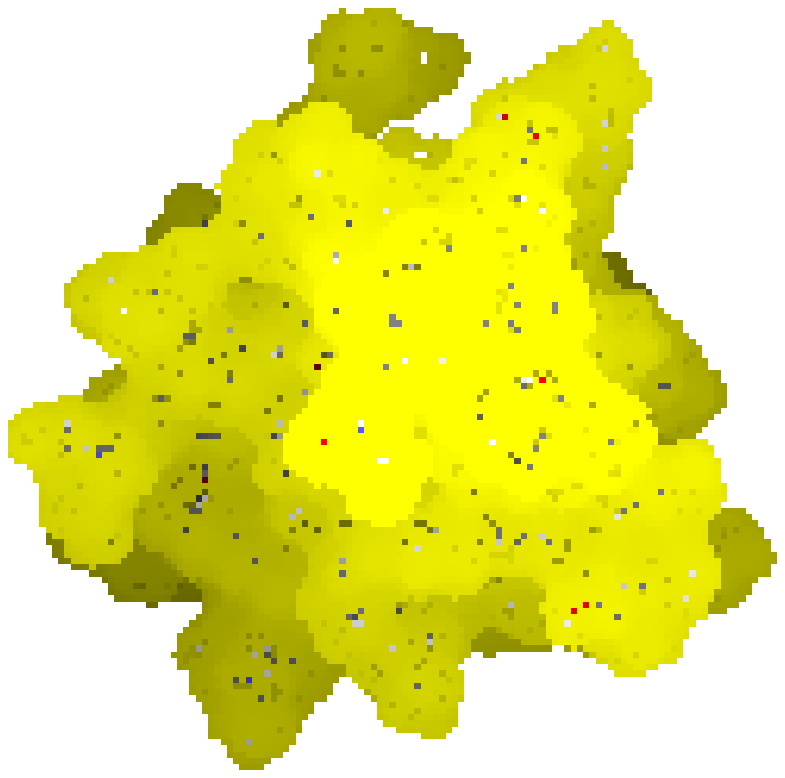}}
\end{minipage}
\begin{minipage}[t]{7.0cm}
\centering
\scalebox{0.5}{\includegraphics{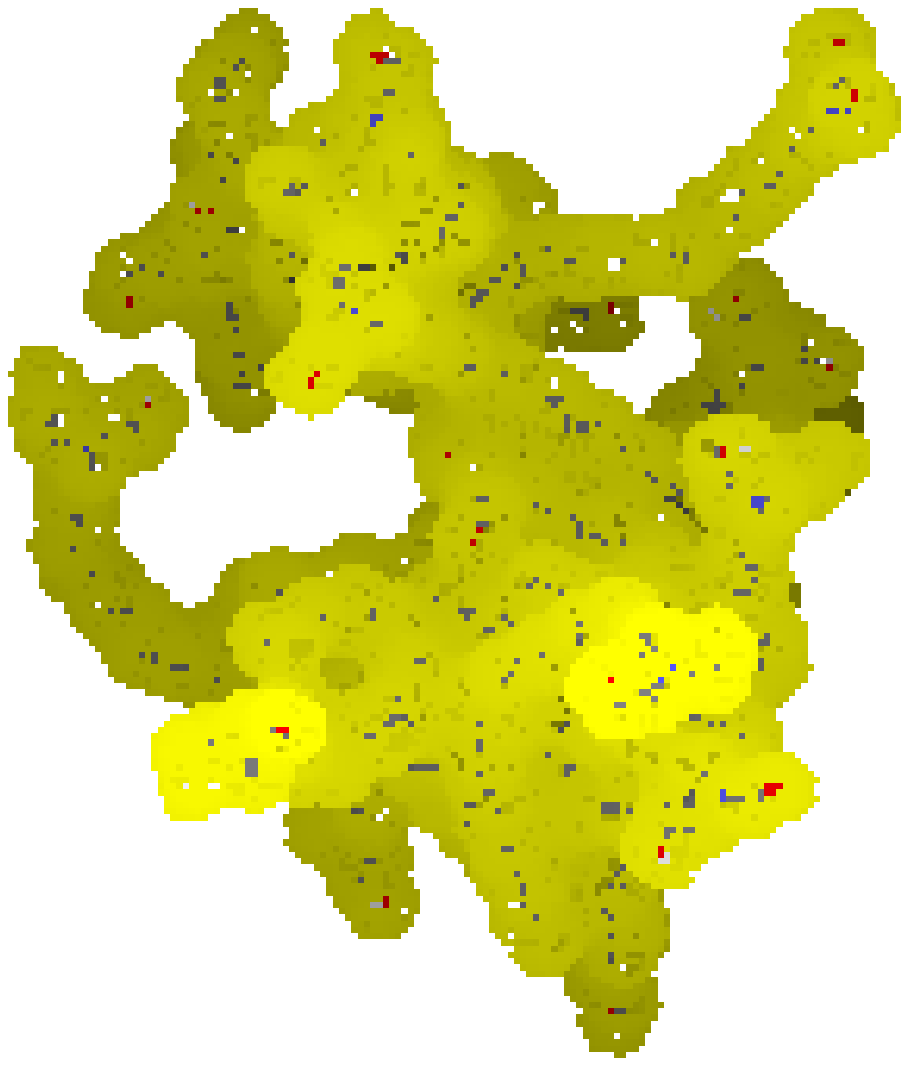}}
\scalebox{0.5}{\includegraphics{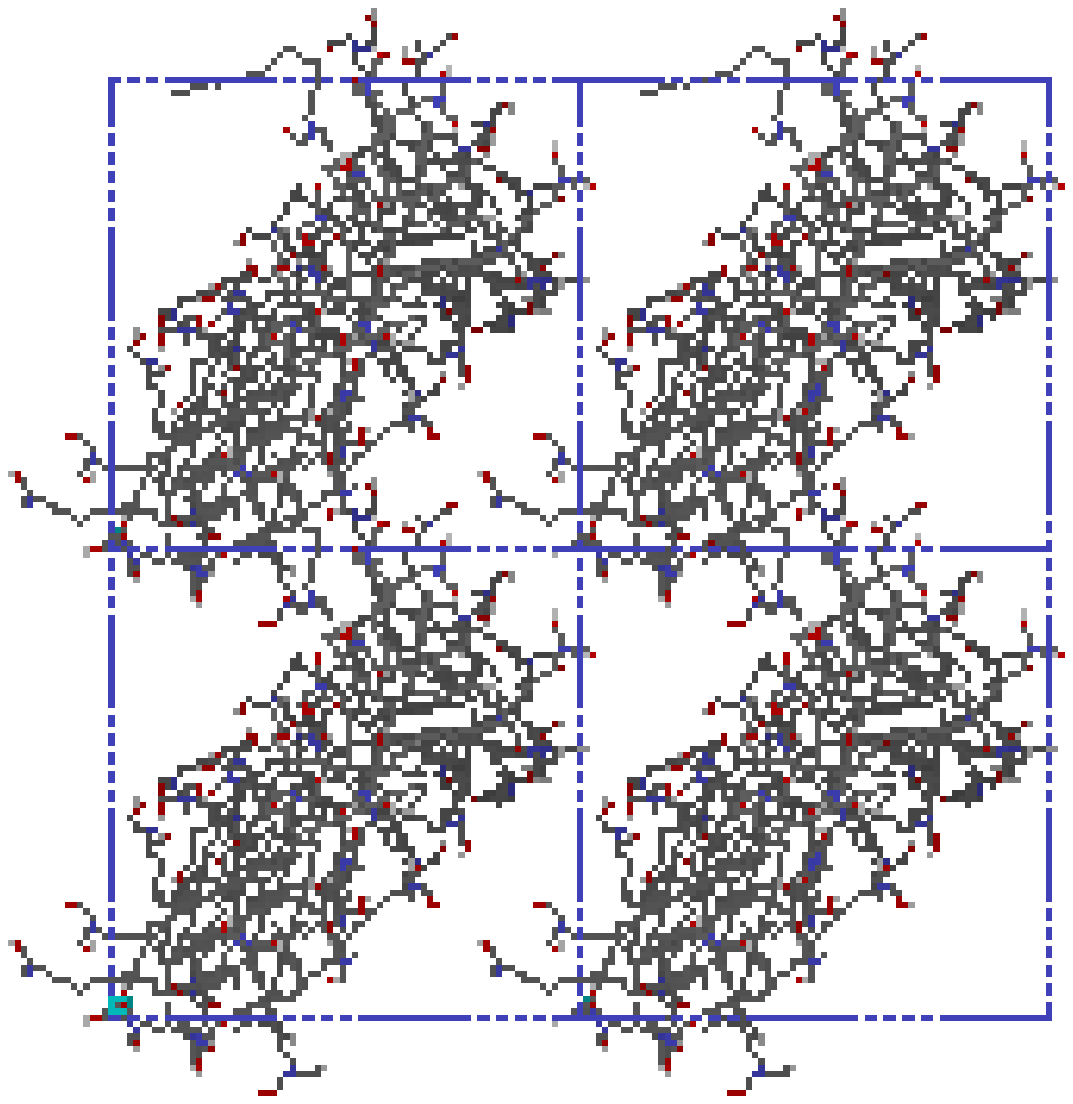}}
\end{minipage}
\end{minipage}
\caption{\textbf{Instantaneous configurations of an initially random
distribution of 50 EMAC molecules at the beginning of a molecular
dynamics simulation
(upper left), after 500 ps for the two resulting micelles (shown
separately at upper right and lower
left in terms of their Connolly surfaces) each containing twenty
monomers and at the end of the simulation (t = 1200 ps) (the 2D
projection of the system shows what appears to be a single rod-like micelle).}}
\label{micelle7}
\end{figure}

A simulation was set up starting from a random distribution of
$50$ EMAC surfactant molecules in a simulation cell containing $3119$ water molecules
and $50$ chloride ions. Figure~\ref{micelle7} displays the initial
configuration of the system from which molecular dynamics was perfomed
for $1.3$ ns. At
t = 500 ps, two surfactant clusters can already be seen, each
containing $20$ molecules; these are
shown in figure~\ref{micelle7}. The Connolly
surface has been
calculated for all atoms of the surfactant molecules and is displayed
in yellow. The first micelle does not exhibit a dense
spherical shape,
as can be seen by the presence of several voids in the structure.
The second micelle is spherical, however, with a uniform density of surfactant
molecules in all directions, leading to a compact 
structure. The structural differences between the two micelles are
summarized succinctly in terms of the various characteristic geometric
parameters reported in
table~\ref{radiusCF}. By the end of the simulation, no significant change
occurs, and the final configuration, consisting of two micelles of
sizes $22$ and $20$, is shown in figure~\ref{micelle7}. One micelle has
grown while the other has kept its size unchanged; eight
monomers remain elsewhere in the solution. The final geometrical
parameters for
the two clusters are also reported in table~\ref{radiusCF}.

\subsection{Infinite cylindrical micelle with electrolyte and
co-surfactant}
\begin{figure}[p]
\centering
\scalebox{0.5}{\includegraphics{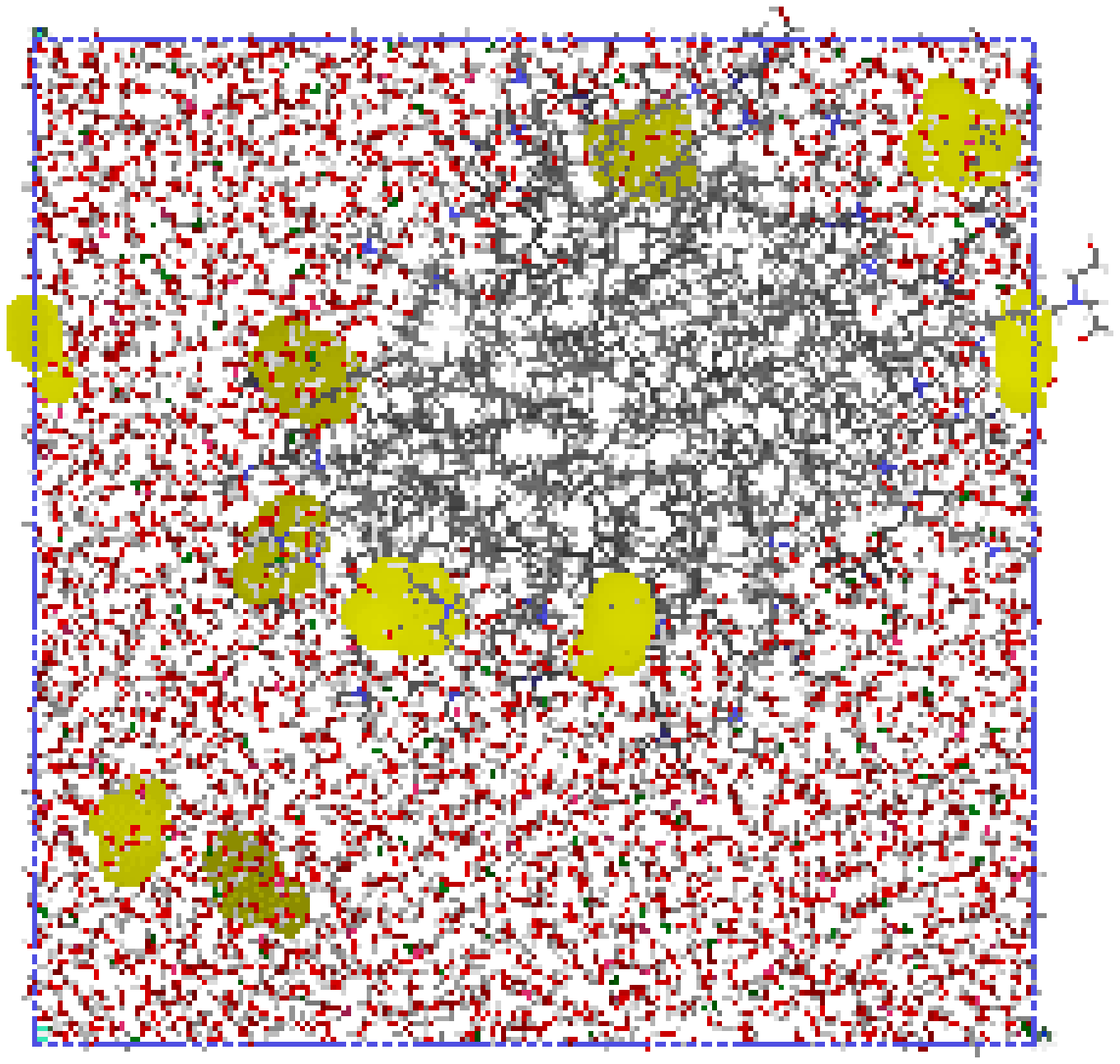}}
\scalebox{0.7}{\includegraphics{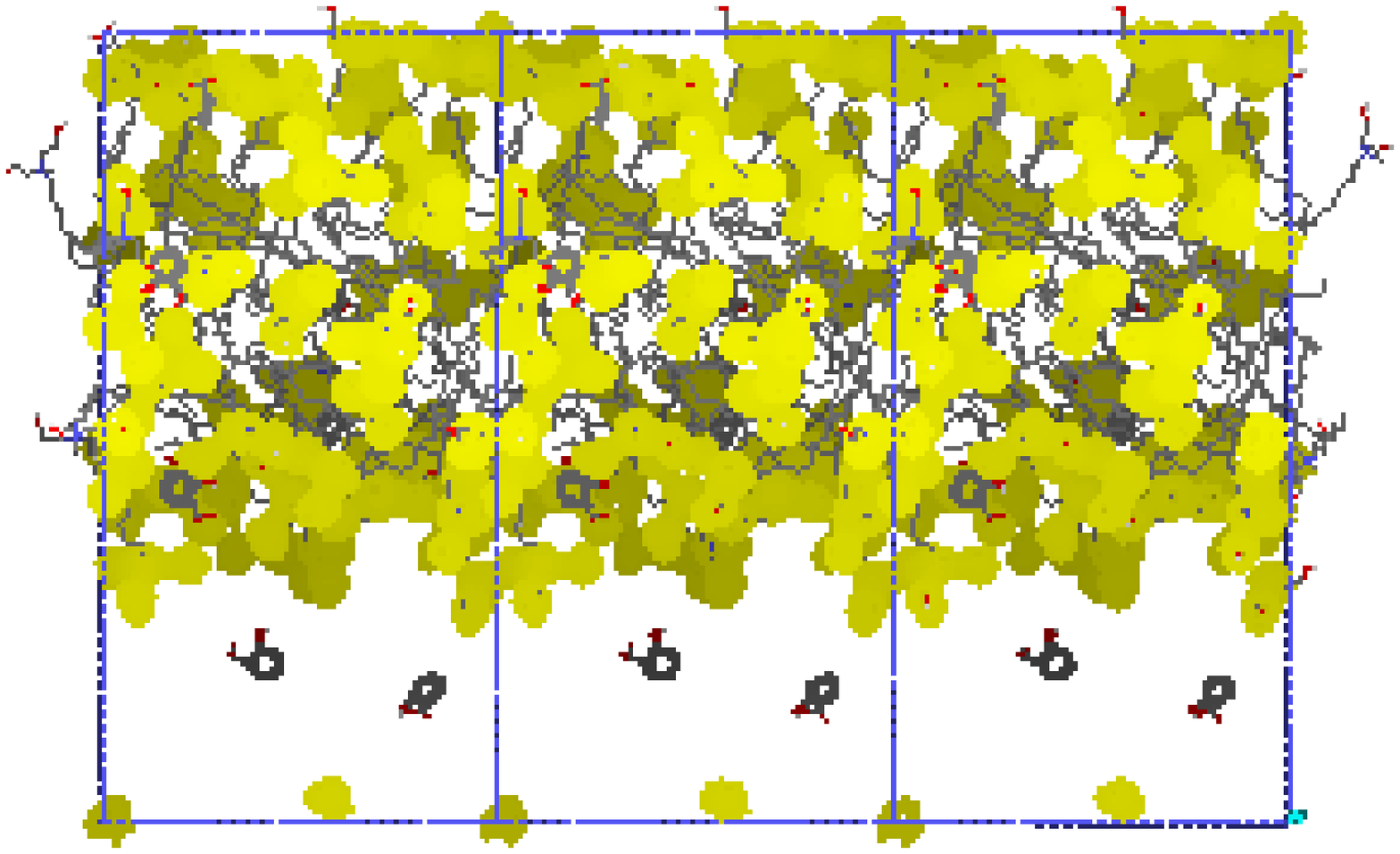}}
\caption{\textbf{Final instantaneous configurations (after 400 ps of a
molecular dynamics simulation) of an initially
infinite EMAC cylindrical micelle with added electrolyte and
salicylate co-surfactant showing views perpendicular (top) and
parallel (bottom) to the major axis of the cylinder. The Connolly
surface is shown for the co-surfactant molecules (top) and for the surfactant
head groups (bottom).}}
\label{micelle8}
\end{figure}

An infinite cylindrical micelle was built as
previously described. It contained $50$ molecules, and was surrounded
by $3322$ water molecules, $150$ chloride ions, $109$ sodium ions, and
$9$ salicylate co-surfactant molecules. The co-surfactant molecules
were placed outside the micelle. Molecular dynamics was performed on
this 3D periodic system for $400$
ps. During this period, the cylindrical shape remained stable (as in
the previous study without additional electrolyte and
co-surfactant, see section~\ref{cylemac}), and its
geometrical parameters are reported in table~\ref{radiusCF}. As in the case
without additional electrolyte, the cross-section of the cylinder is
not circular but
rather ovoid, one principal axis being markedly greater than the other.
Figure~\ref{micelle8} shows two views of the micelle at the end
of the simulation. In the first one, all atoms in the system are
displayed. The Connolly surface has been calculated for the co-surfactant
molecules and is displayed in yellow. We can see -as in the case
without added salt-that no water molecules have penetrated the
micelle. A large number (six out of nine) of co-surfactant molecules have
entered the micelle structure, being adsorbed on its
surface and surrounded by the surfactant head groups. The second
snapshot shows the micelle in a view
parallel to the axis of the cylinder; the Connolly surface has been
calculated for the atoms of the head groups only. The surfactant
molecules seem to be homogeneously dispersed along the structure, and
there is no appearance of weak points, or an incipient site of fragmentation.
Nevertheless, we can see that
large voids remain in the structure, providing evidence of direct
contacts between the hydrophobic  tails of the surfactant and the water
molecules at the micellar surface. These voids are not associated with
inhomogeneous density regions but are more likely due to an insufficient
initial density of surfactant and co-surfactant.

\section{Discussion of results}
\label{discussion}
\subsection{$\rm C_{9}TAC$ surfactant micelles}
\begin{figure}[p]
\centering
\scalebox{0.5}{\includegraphics{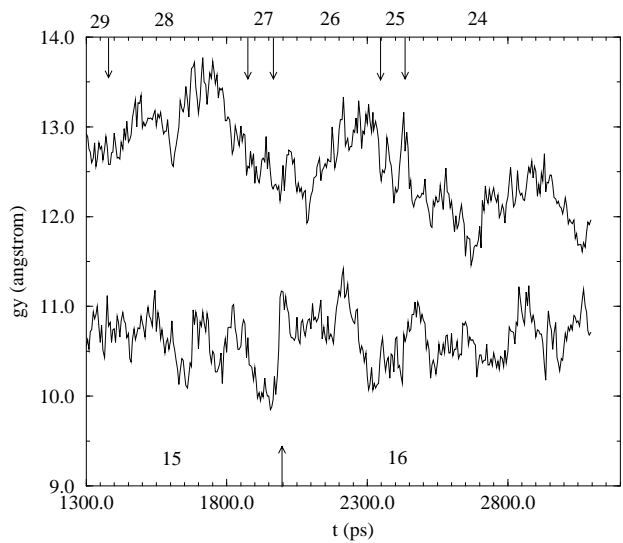}}
\caption{\textbf{Evolution of the radius of
gyration for the larger (upper curve) and the smaller (lower curve) $\rm
\mathbf C_{9}TAC$ micelle. The data are reported only beyond the initial 1.3 ns of
a molecular dynamics simulation. The initial configuration was a
spherical micelle (see figure~\ref{micelle1}). The arrows on the upper
curve designate instants at which individual surfactant monomers leave
the pulsating micellar cluster. The integers between the arrows refer
to the aggregation number of each micelle during the simulation.}}
\label{gyration}
\end{figure}

\begin{figure}[p]
\centering
\scalebox{0.5}{\includegraphics{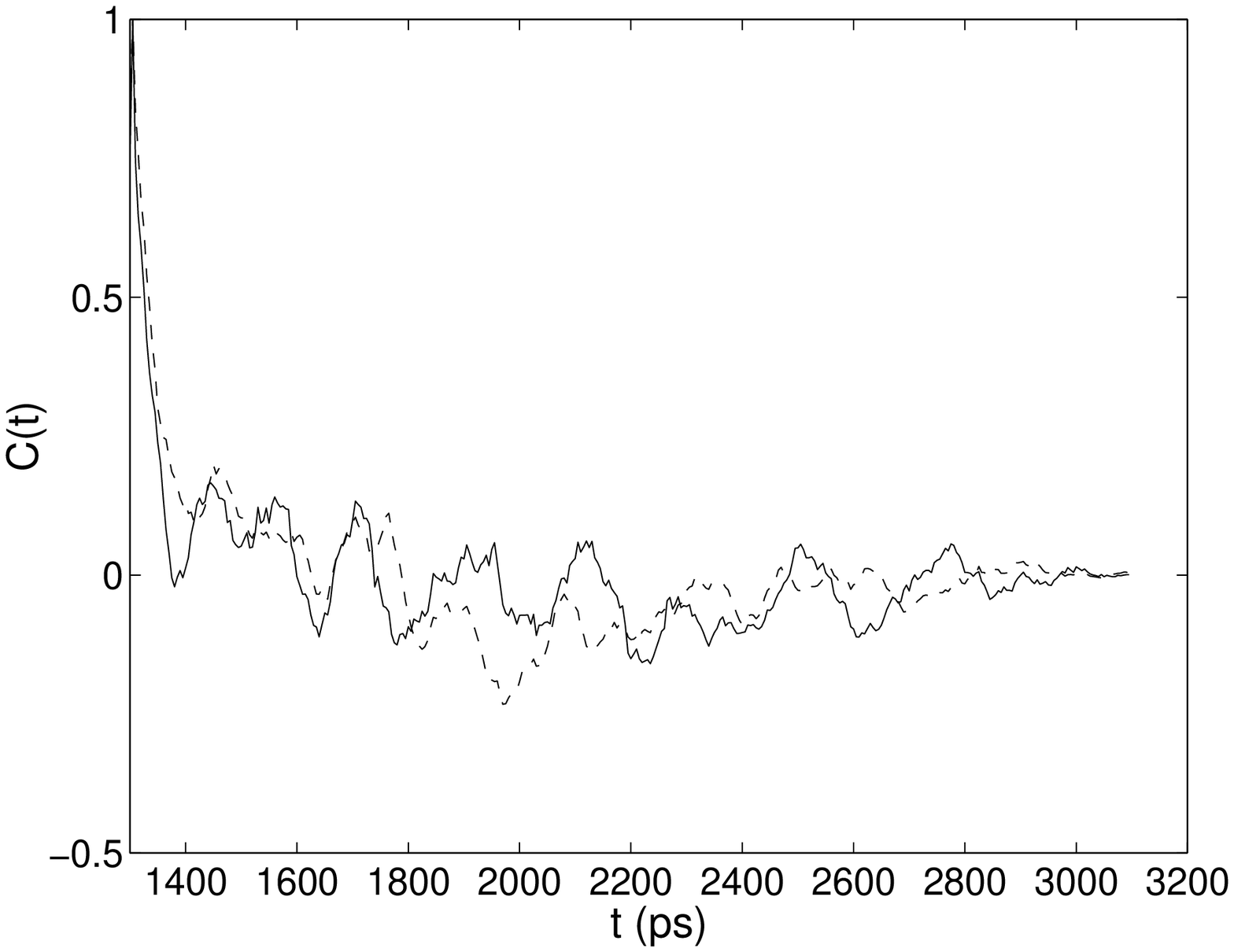}}
\caption{\textbf{Evolution of the autocorrelation function of the ratios of the
length of the principal axes of the smaller micelle (described in
figures~\ref{micelle1},~\ref{gyration}). The continuous and dotted
curves are the autocorrelation functions of the ratio $\frac{I_{2}}{I_{3}}$ and
$\frac{I_{1}}{I_{3}}$ respectively.}}
\label{corelsmall}
\end{figure}

\begin{figure}[p]
\centering
\scalebox{0.5}{\includegraphics{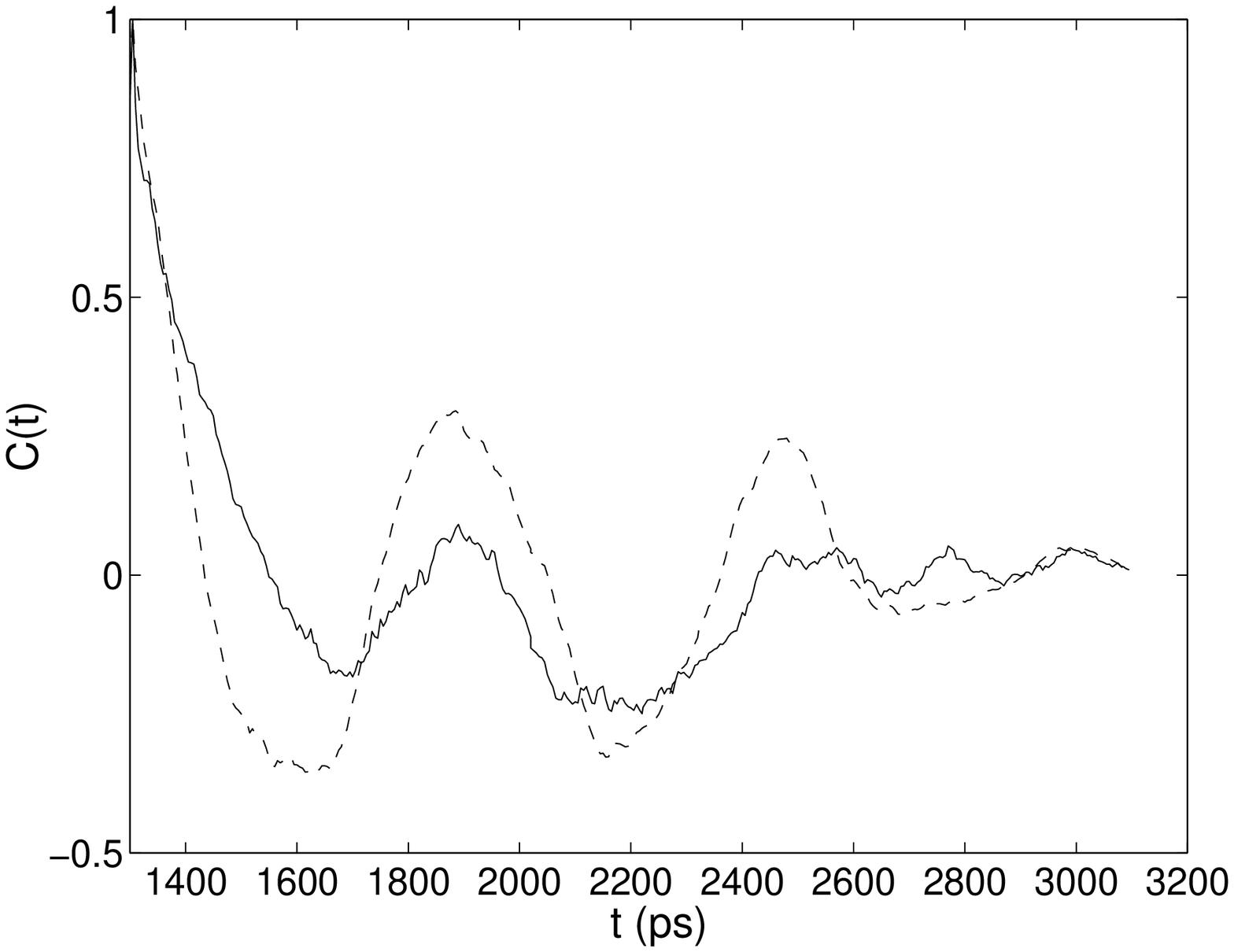}}
\caption{\textbf{Evolution of the autocorrelation function of the ratio of the
length of the principal axes of the larger micelle (described in
figures~\ref{micelle1},~\ref{gyration}). The continuous and dotted
curves are the autocorrelation functions of the ratio $\frac{I_{2}}{I_{3}}$ and
$\frac{I_{1}}{I_{3}}$ respectively.}}
\label{corelbig}
\end{figure}

\begin{figure}[p]
\centering
\scalebox{0.5}{\includegraphics{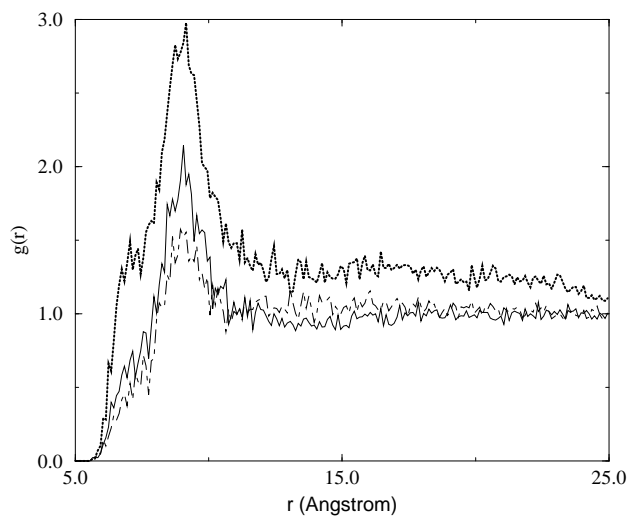}}
\caption{\textbf{Radial distribution function for $\rm \mathbf C_{9}TAC$
surfactant between nitrogen
atoms, computed by averaging over the last
nanosecond of molecular dynamics simulation for an initially spherical
micelle (solid line), a cylindrical micelle
with added electrolyte (dotted line), and an initially random distribution
of monomers (dot-dashed line). Data are taken from the MD
simulations shown in figures~\ref{micelle1},~\ref{micelle3},~\ref{micelle4}.}}
\label{TACgrNN}
\end{figure}

\begin{figure}[p]
\centering
\scalebox{0.5}{\includegraphics{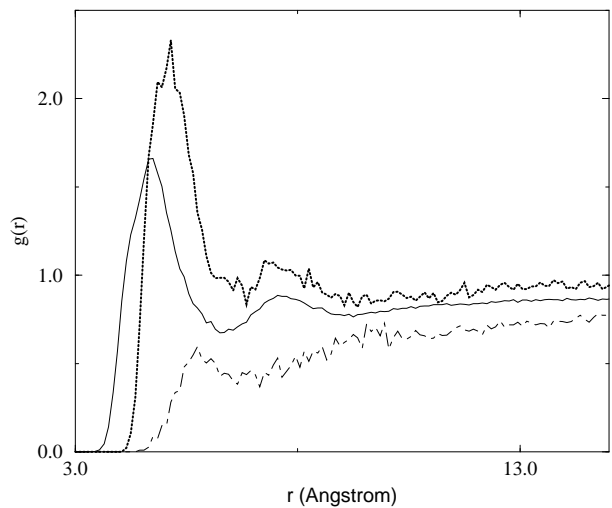}}
\caption{\textbf{Radial distribution function for $\rm \mathbf C_{9}TAC$
surfactant between nitrogen and
oxygen atoms (solid line), nitrogen atoms and chloride ions (dotted line)
and nitrogen atoms and sodium ions (dotted-dashed line) computed by
averaging over the last
nanosecond of molecular dynamics simulation for the initially
cylindrical micelle with added electrolyte (see figure~\ref{micelle3}).}}
\label{TACgrN-other}
\end{figure}

For micellar systems, it appears that the characteristic time for full
thermodynamic equilibration is generally
greater than the microsecond scale. Nevertheless, the
study of the behaviour of such systems on nanosecond timescales can
provide useful information on the dynamics and structure of
micelles. Indeed, some processes like monomer absorption or
desorption, or the fragmentation of cylindrical or spherical micelles,
can be simulated on this timescale. 
Starting from different initial configurations (spherical, cylindrical or
random distribution of surfactant molecules) of essentially similar
composition, we observe that
these systems tend to evolve to the same state, which is
characterised by the existence of two quasi-spherical micellar clusters of
small aggregation number (between $15$ and $20$)
and a few monomers isolated in the water. The radius of
gyration of each micelle is roughly equal to $11$ $\rm \AA$, and the
radii of the micelles (calculated as the averaged distance between the
centre of the micelle and the nitrogen atoms) is $7.5 - 8$ $\rm
\AA$. These micelles are non spherical \emph{on average}, the average ratio of
the length of the principal axes being significantly different from
unity.\\
The way
this final state is reached starting from distinct initial
configurations is different: in the case of a spherical or cylindrical
micelle, we observe fragmentation into two smaller clusters. In
the case of a random distribution of surfactant cations, we directly observe
the \emph{formation} of micelles. This process starts with the
appearance of a disordered aggregate of surfactant molecules, of small
size. A reorganisation of the
surfactant molecules occurs, the head groups becoming anchored at the
micellar surface. The time needed for some kinds of intramicellar
rearrangements (following fragmentation) can be very small. As an
illustration,  a simulation was performed to investigate the
rearrangements of surfactant molecules at the end of a finite rodlike
micelle. The initial configuration was
a finite rodlike micelle lacking end-caps, thus exposing extensive
amounts of its hydrophobic interior to direct contact with water. In the
first $10$ ps of simulation, the surfactant molecules move to create
hemispherical end-caps. The time needed for several molecules to
rearrange inside a micelle is thus very short, at least when high
energy configurations are involved.\\
In general, growth of micelles is achieved by the
absorption of groups (typically dimers or trimers) of surfactant cations rather
than by stepwise adsorption of monomers. We believe that this mechanism
of micelle growth, corresponding to a
Smoluchowski kinetic model:
\begin{equation}
C_{r} + C_{s}  \begin{array}{c} \stackrel{\stackrel{\textstyle a_{r,s}}{\textstyle \rightharpoonup}}
{\stackrel{\textstyle \leftharpoondown}{\textstyle b_{r+s}}} \end{array} C_{r+s}
\label{smol}
\end{equation}
where $C_{r}$ is a micellar cluster of aggregation number $r$ and
$a_{r,s}$ and $b_{r+s}$ are the forward and backward rate coefficients
for aggregation and fragmentation processes,
is valid when the system is far from
equilibrium and at surfactant concentrations well above the c.m.c.,
because it leads to a faster rate to reach equilibrium than with a
Becker-D\"{o}ring scheme:
\begin{equation}
C_{r} + C  \begin{array}{c} \stackrel{\stackrel{\textstyle a_{r}}{\textstyle \rightharpoonup}}
{\stackrel{\textstyle \leftharpoondown}{\textstyle b_{r+1}}} \end{array} C_{r+1}
\end{equation}
i.e. a purely stepwise addition or removal of surfactant monomer~\cite{volmer3}.
On the other hand, starting from a spherical or cylindrical
micelle, we can observe its break up. In both cases, the initial micelle
breaks into two clusters of different size, one containing
approximately $15$
molecules and the other one $30$ molecules. The fact that a
fragmentation process is observed confirms the validity of the
Smoluchowski mechanism when the system is far from equilibrium.
The size of the smaller
micelle corresponds approximately to the equilibrium size of a $\rm
C_{9}TAC$ micelle, which remains essentially unchanged for the rest of the
simulation. The size of the larger micelle does not correspond
to the equilibrium size, and we observe a decrease of its
aggregation number with time. This decrease is a stepwise process,
following the Becker-D\"{o}ring scheme. 
It is observed that the initial
micelle produces two micelles of different sizes (one of which appears
to be closer to the mean equilibrium cluster size) rather than
producing two micelles of equal
size, but larger than the equilibrium size.
For systems not far from equilibrium, and/or for surfactant
concentrations not greatly exceeding the c.m.c., a Becker-D\"{o}ring
scheme is more likely to be observed~\cite{volmer3}, characterized by a stepwise change in the
aggregation number.\\
Figure~\ref{gyration} shows the evolution of the radius of gyration of
the two micelles (initially one spherical micelle) during the last 2
ns of simulation. The lower curve is associated with the smaller
micelle which grows from $15$ to $16$ molecules during this part of
the simulation. The radius of gyration of this micelle
fluctuates around its average value of $10.63$ $\rm \AA$, with a
standard deviation equal to $0.29$. This may correspond to
the behaviour of a micelle near equilibrium. The upper curve is
associated with the bigger
micelle and arrows indicate the moments at which individual surfactant monomers
leave the micelle. The main feature exhibited by the radius of gyration is a
tendency to decrease,
and this decrease is associated with the loss of monomers.
Moreover, we can see that the curve exhibits oscillations of large
amplitude (greater than a typical fluctuation). This
corresponds in fact to an
expansion-contraction process. The expansion of the micelle
corresponds to an elongation in one direction; the surfactant
molecules leave the micelle during the contraction process. This
contraction indeed induces an increase in the repulsive interactions
between head groups, leading to the desorption of one molecule.\\
In order to highlight these shape fluctuation phenomena in
the two micelles, the autocorrelation functions of fluctuations in the ratio of the
lengths of the principal axes have been computed \mbox{as :}
\begin{equation}
C(t)=\frac{\langle \delta R(t) \delta R(0) \rangle}{\langle \delta R(0) \delta R(0) \rangle}
\end{equation}
where $\delta R(t)$, defined in~\cite{watanabe}, is:
\begin{equation}
\delta R(t) = R(t) - \langle R(t) \rangle,
\end{equation}
and $R(t)$ is the ratio of the lengths of the principal axes.
These are displayed in
figures~\ref{corelsmall} and \ref{corelbig}. The curve pertaining to
the smaller cluster exhibits
quasi-periodic oscillations, suggesting a time scale for the shape
fluctuations of about $50$ ps. This agrees well with what has
been previously observed in a 100 ps molecular dynamics simulation of
a sodium octanoate micelle containing $15$ surfactant anions in
water~\cite{watanabe}, where the time
scale for a shape fluctuation was found to be equal to 30 ps.\\
The curve associated with
the larger micelle (figure~\ref{corelbig}) shows a different behaviour:
periodic oscillations of large amplitude are seen, with a
periodicity of $\sim 500$ ps. This time interval is associated with the slow
expansion-contraction process described above. In this system, the fast shape
fluctuation process has disappeared.
The quasi-periodicity observed in these oscillations is possibly due
to the fact that the rate of monomer desorption varies slowly with the
cluster size. \\
These results indicate that a shape fluctuation can be coupled with
a desorption process, leading to a considerable increase in its
characteristic time scale. The time scale of shape fluctuations in
spherical micelles thus depends on their sizes (and therefore on the
proximity to equilibrium).\\
When electrolyte is added to the system, the dynamic behaviour is changed:
the initially infinite micelle breaks up, loosing its infinite length. It thus
becomes a finite rod, but still retains cylindrical symmetry. This
small rod-like micelle is stable over the $2.5$ ns of simulation.
The addition of salt is known to reduce the effective surface area per
head group~\cite{israelachvili,israel2}, leading to the stabilization
of the cylindrical shape. The
head groups, in a rod-like micelle, are thus expected to be closer
together
than in a spherical micelle. Figure~\ref{TACgrNN} shows the radial
distribution function between nitrogen atoms, calculated over the last
nanosecond of the simulations described in section~\ref{restac}, in
the three cases of a
spherical micelle, a cylindrical micelle, and an initially random
configuration. The three curves exhibit their main peak at roughly the same
distance ($9\rm \AA$), while some differences can be discerned at shorter
distances.
The curves pertaining to the spherical micelle and the
initially random surfactant configuration are very similar.
Calculations of the mean total
energies and volumes of the simulation cells confirm the similarity between
these two systems. By contrast, the curve associated with the
cylindrical micelle exhibits a more pronounced secondary peak at short
N-N distances, providing
evidence of a large number of close head-group contacts for this
micellar structure.
Figure~\ref{TACgrN-other} shows the radial distribution functions
calculated between nitrogen atoms and respectively chloride ions,
sodium ions, and
oxygen atoms of water molecules. The short distances of the N-O (water)
and N-Cl radial distribution functions indicate a high degree of
structuring around the polar head groups. The closest distances
correspond to nitrogen-oxygen interactions. 
Moreover, the appearance of a second (outer) peak in the N-Cl
and N-O (water) radial distribution functions is associated with the
existence of a second solvation shell surrounding the polar head
group. This second peak in these curves is at the same distance
in both cases. The sodium cations are located between the two solvation
shells, but remain closer to the inner one.

\subsection{EMAC micelles}
\begin{figure}[p]
\centering
\scalebox{0.5}{\includegraphics{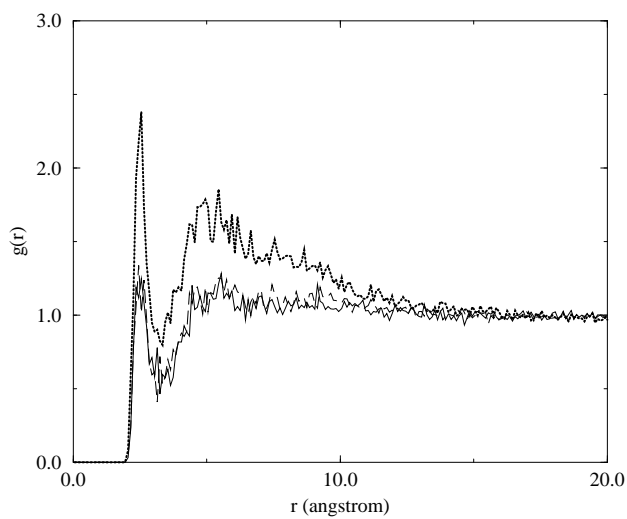}}
\caption{\textbf{Radial distribution function between the hydrogen atoms of
the hydroxyl group of the EMAC surfactant molecule and the chloride ions,
computed by averaging over the last nanosecond of a molecular dynamics
simulation for an initially
spherical micelle (solid line), an infinite cylindrical micelle
(dotted line) and an initially random configuration of monomers (
dashed line). The data are taken from MD simulations shown in
figures~\ref{micelle5},~\ref{micelle6},~\ref{micelle7}.}}
\label{CFgrHoCl}
\end{figure}

\begin{figure}[p]
\centering
\scalebox{0.5}{\includegraphics{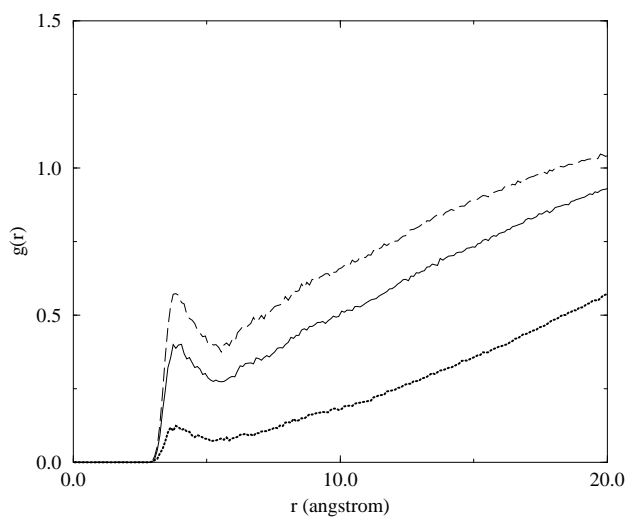}}
\caption{\textbf{Radial distribution function
between the carbon atoms of
the terminal methyl group of EMAC and the oxygen atoms of water
computed by averaging
over the last nanosecond of a molecular dynamics simulation for an initially
spherical micelle (solid line), an infinite cylindrical micelle
(dotted line) and
an initially random configuration of monomers (dashed line). The
data are taken from MD simulations shown in
figures~\ref{micelle5},~\ref{micelle6},~\ref{micelle7}.}}
\label{CFgrC3jbOtip}
\end{figure}

\begin{figure}[p]
\centering
\scalebox{0.5}{\includegraphics{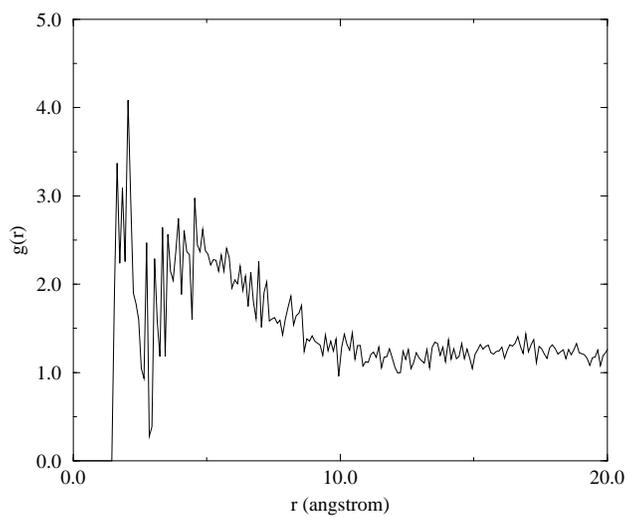}}
\caption{\textbf{Radial distribution function between the hydrogen
atoms of
the hydroxyl group of EMAC and the oxygen atoms of
the salicylate co-surfactant molecule computed by averaging over 300
ps of a molecular dynamics simulation for an initially
infinite cylindrical EMAC micelle with added electrolyte and co-surfactant
(see figure~\ref{micelle8}).}}
\label{CFgrHoOnewstart12}
\end{figure}

As in the case of $\rm C_{9}TAC$ micelles, various starting
configurations were used to
investigate the dynamical behaviour of EMAC micelles. In the absence of
electrolyte or co-surfactant, this molecule is known to form
spherical micelles~\cite{hughes}.
Starting either from a
spherical or a cylindrical micelle, the system was found to keep its
initial shape
over a few nanoseconds of molecular dynamics simulation. In the case
of wormlike micelles, although some evidence of incipient
fragmentation was detected,
the simulations were not performed over a large enough time to confirm this.
It was also found that
cross-sections of the worm were not circular, but rather elliptical
as in the case of ellipsoidal (initially spherical)
sodium octanoate micelles~\cite{kuhn2}. The dynamics of
EMAC aggregation and fragmentation processes is found to be slower
than those for
$\rm C_{9}TAC$ self-assembly. This might be interpreted in terms of the length of
the hydrophobic tail of the surfactant (from steric and energetic
considerations, a long hydrophobic tail in a liquid would be more difficult to
displace than a short one, and its diffusion coefficient is smaller).
Nevertheless, we can see some differences between the spherical and
the cylindrical micelles of EMAC molecules in comparison with $\rm C_{9}TAC$.
Figure~\ref{CFgrHoCl} displays the radial
distribution function between the hydrogen atoms of the hydroxyl group
of the EMAC surfactant cation and the chloride anions. The properties
of this function pertaining
to the spherical micelle and the initially random distribution of
surfactant cations are
similar. There is a close contact between these two atoms at about $2.5$
$\rm \AA$, followed by an exclusion domain. The close proximity
is indicative of a strong association between these atoms; while
the exclusion domain is due to the electrostatic repulsion between
chloride anions.
Beyond the exclusion domain, the probability of finding a chloride ion
reaches a value equal to that for the bulk
concentration of electrolyte. In the case of the wormlike micelle, the
first peak is higher than those corresponding to other shapes, and a second peak is
clearly present. The surfactant head groups within EMAC in a wormlike
micelle are thus seen to exhibit a stronger
interaction with the counterions than in the case of a spherical
micelle.\\
Figure~\ref{CFgrC3jbOtip} displays the radial distribution function
between the terminal methyl group and the oxygen atoms of water. We can
see that in the three cases, there is a first peak at about $4$ $\rm \AA$.
In the case of the cylindrical micelle, it will be recalled
(see figure~\ref{micelle6})
that there is no water penetration inside the micelle. The short
C-O contact distance seen in the radial distribution function displayed in
figure~\ref{CFgrC3jbOtip} is thus due to the
fact that the terminal methyl group is not located at the centre of the
micelle, but rather is in direct contact with surrounding
water.
We can see that the location of the first peak and the probability of finding a short
contact between the terminal methyl group and the water molecule
both increase when going from the cylindrical to the spherical micelle,
and finally for the curve resulting from the initially random configuration
of EMAC monomers.
These trends can be interpreted in terms of the effective surface area per head group,
and the compactness of the resulting structure.
In micelles with a cylindrical geometry, the head groups are closer to
each other than
they are in a spherical
micelle~\cite{israelachvili,israel2}, so that the compactness of the
former structure is greater, leading to
reduced water penetration. 
Moreover, in the case of the initially
random monomer distribution, the two resulting micellar clusters
contain half the number of molecules as in the case of the spherical
micelle studied here, while their radii (see table~\ref{radiusCF}) are
roughly identical to those of the spherical cluster. The structure of
the two small clusters is thus less compact, leading to a larger
probability of close contacts between the hydrophobic chain and the
water molecules.
Even if there are some differences concerning the
water penetration process in $\rm C_{9}TAC$ and EMAC micelles, in
all cases we have observed that the core of the
micelle is completely impervious to water.\\
In the case of the cylindrical micelle with added electrolyte and
salicylate
co-surfactant, the radial distribution function between the hydrogen
atoms of the hydroxyl group of the surfactant molecule and the oxygen
atoms of the co-surfactant molecule is displayed in
figure~\ref{CFgrHoOnewstart12}. We can see that there is a first peak
at about $2$ $\rm \AA$, providing evidence of hydrogen bonding between these
two species. As was also noticed earlier (see figure~\ref{micelle8}), a large
number of co-surfactant anions are integrated into the surface of the micelle.
These anions do not penetrate the core of the micelle but rather
remain adsorbed on its surface, enjoying a strong interaction with the
surfactant
head groups. This interaction decreases the effective surface
area per head group, and enhances the formation of rod-like micelles,
precisely as proposed in simple geometrical
theories~\cite{israelachvili,israel2}.
The existence of hydrogen bonds between the head groups of the
surfactant and the solvent is evidently not a necessary
condition for the
formation of micelles~\cite{smit}. Nevertheless, in this particular
systems, it seems to play a
key role in the attachment of co-surfactant molecules to the micelle.
We have also observed the presence of hydrogen bonding between the
hydrogen atoms of the hydroxyl group of the EMAC surfactant and the
oxygen atoms of water molecules. However, there is no evidence
of hydrogen bonding involving the oxygen atom of the hydroxyl group of
the EMAC surfactant and the hydrogen atom of water molecules nor
between two hydroxyl groups of different surfactant head groups. No
interaction between two salicylate molecules has been seen.

\section{Conclusions}
Large-scale molecular dynamics simulations have been performed to investigate
the structural and dynamical properties of self-assembled cationic
surfactants in aqueous solution. One of the surfactants was
comparatively short-chained ($\rm C_{9}TAC$), the other long-chained
(EMAC). The nanosecond regime has been reached, allowing the
study of the dynamics of various self-assembly processes (growth and
fragmentation of micelles, surfactant monomer insertion or removal).
The mechanism of micelle formation at the molecular level has been
described. We have interpreted separately the kinetics of these dynamical
processes when the system is variously far from or near equilibrium. In
the former case, a Smoluchowski-type scheme is obeyed, according to which
micelles coalesce or fragment. In the latter case, a Becker-D\"{o}ring
scheme is observed~\cite{volmer3}, where only step-by-step monomer exchanges take place.
It was also found that, for an oversized micellar cluster, the
step-by-step elimination of surfactant monomers is associated with a
slow
expansion-contraction process of the micelle, with a characteristic
time period of $500$ ps. On the
other hand, a characteristic time period of $50$ ps for shape fluctuations
was found in the case of a spherical $\rm C_{9}TAC$ micelle with size
close to the mean cluster size.
The dynamics of the EMAC molecule is slower,
and the equilibrium state has not been reached starting from different
configurations. This difference in the time scales of the dynamics
between the $\rm C_{9}TAC$ and the EMAC molecule is attributed mainly
to the size difference of the two cationic surfactant molecules. The effect
of co-surfactant has been investigated and
hydrogen-bonding with the head groups of the surfactant molecule has
been found to play an important and probably a key role in stabilizing
the wormlike assembly
of EMAC cations. Finally, we have found that the
penetration of water molecules inside micelles is not important in
any instance examined, at least over time scales up to a few
nanoseconds.

\section{Acknowledgements}
This work has been done in collaboration with Silicon Graphics Inc.
who have provided access to a number of large parallel machines.
Daron Green and John Carpenter are gratefully acknowledged for their
generous technical assistance.
Fruitful discussions with Trevor Hughes and Edo Boek are gratefully
acknowledged. Mike Stapleton, Andreas Bick and Richard Painter of
Molecular Simulations Inc. are also thanked for their support of this work.

\newpage

\begin{table}[p]
\begin{center}
\caption{\textbf{Geometry (interatomic distance and angle) and
intermolecular potential parameters (van der Waals parameters and
coulombic charges on atoms) of the TIP3P water molecule.}}
\label{waterpot}
\begin{tabular}{cccccc}
$\emph{r}$(OH), \AA & $\angle$HOH,deg & $\sigma_{oo}$, \AA & $\epsilon_{oo}$,cal.mo$l^{-1}$ &
q(O) & q(H)\\
\hline
0.9572 &104.52 &3.15 &152.07 &-0.834 & 0.417
\end{tabular}
\end{center}
\end{table}

\begin{table}[h]
\begin{center}
\caption{\textbf{Bulk density $\rho$ and diffusion coefficient $D$ of
water at 300 K. Comparison between the calculated values obtained
using the TIP3P force-field and the experimental data.}}
\label{waterprop}
\begin{tabular}{ccc}
 &$\rho (g/cm^{3})$ & $D(10^{5} cm^{2}/s)$ \\
\hline
TIP3P & 1.02 & 7.9\\
exp. & 1.0 & 2.4
\end{tabular}
\end{center}
\end{table}

\begin{table}[h]
\begin{center}
\caption{\textbf{Calculated values of the radius of gyration $\rm
R_{G}$, the ratios of the lengths of the principal axes, and the radius
of the micelle $R_{N}$ (calculated as the distance from the centre of
the micelle to the nitrogen atoms), for $\rm \mathbf C_{9}TAC$
micelles. (1) and (2) refer respectively to the smaller and
the larger micelles when two micelles are present simultaneously. The
first letter in the first column corresponds either to ``upper'' (U) or
``lower'' (L) and the second one to ``left'' (L) or ``right'' (R). ND
stands for ``not displayed''.}}
\label{radiusTAC}
\begin{tabular}{cccccccc}
figure & initial shape & time (ps) & $\rm R_{G}$ $\rm (\AA)$ & $\frac{I_{1}}{I_{2}}$  & $\frac{I_{2}}{I_{3}}$  &
$\frac{I_{1}}{I_{3}}$ & $\rm R_{N}$ $\rm (\AA)$ \\
\hline
\vspace{1.5mm}
fig~\ref{micelle1} - UL & spherical & 50 & 14.85  & 1.05 & 1.10 & 1.15 & 10.26\\
\vspace{1.5mm}
fig~\ref{micelle1} - UR & spherical & 600 & 15.95 & 0.75 & 1.81 & 1.36 & 10.59\\
fig~\ref{micelle1} - LL & spherical & 1100 (1)& 10.63 & 1.11 & 0.60 & 0.67 & 7.44 \\
\vspace{1.5mm}
fig~\ref{micelle1} - LL & spherical & 1100 (2) & 12.46 & 1.08 & 0.90 & 0.97 & 8.77 \\
fig~\ref{micelle1} - LR & spherical & 3100 (1)& 10.76 & 1.21 & 0.74 & 0.90 & 7.59 \\
\vspace{3mm}
fig~\ref{micelle1} - LR & spherical & 3100 (2)& 12.07 & 0.91 & 1.19 & 1.08 & 8.39 \\
\vspace{1.5mm}
fig~\ref{micelle2} - LL & cylindrical & 700 & 10.40  & 1.43 & 0.75 & 1.07 & 7.40 \\
fig~\ref{micelle2} - ND & cylindrical & 1100 (1)& 10.67 & 0.76 & 0.88 & 0.67 & 7.55\\
\vspace{3mm}
fig~\ref{micelle2} - ND & cylindrical & 1100 (2)& 14.30 & 1.61 & 1.19 & 1.92 & 9.26\\
\vspace{1.5mm}
fig~\ref{micelle3} - UR & cyl + salt & 100 & 15.77  & 1.21 & 1.38 & 1.66 & 10.46\\
\vspace{3mm}
fig~\ref{micelle3} - LL + LR & cyl + salt & 2500 & 15.58 & 1.52 & 1.22 & 1.85 & 10.27\\
fig~\ref{micelle4} - LR & random & 900 (1) & 10.71 & 0.65 & 1.13 & 0.74 & 7.53 \\
fig~\ref{micelle4} - LR & random & 900 (2) & 11.05 & 1.18 & 1.07 & 1.26 & 7.91
\end{tabular}
\end{center}
\end{table}

\begin{table}[h]
\begin{center}
\caption{\textbf{Calculated values of the radius of gyration $\rm
R_{G}$, the ratios of the lengths of the principal axes $I_{1}$,
$I_{2}$, $I_{3}$, and the radii
of the micelle $\rm R_{O}$,  $\rm R_{N}$,  $\rm R_{C22}$,  $\rm
R_{C1}$, taken as the distance between the centre of
the micelle and the oxygen, nitrogen, terminal carbon of the headgroup alkyl
chain, and first carbon of the hydrophobic alkyl tail respectively.
The initial
configurations are a spherical micelle, a cylindrical micelle with or
without electrolyte and co-surfactant, and a
random distribution of EMAC monomers corresponding to the snapshots
shown in
figures~\ref{micelle5},~\ref{micelle6},~\ref{micelle7},~\ref{micelle8}.
(1) and (2) refer respectively to the smaller and
the larger micelles when two micelles are present at the same time.}}
\label{radiusCF}
\begin{tabular}{cccccccccc}
initial conf. & time (ps) & $\rm R_{G}$ $\rm (\AA)$ & $\frac{I_{1}}{I_{2}}$  & $\frac{I_{2}}{I_{3}}$  &
$\frac{I_{1}}{I_{3}}$ & $\rm R_{O}$ $\rm (\AA)$ & $\rm R_{N}$ $\rm (\AA)$ & $\rm
R_{C22}$ $\rm (\AA)$ & $\rm R_{C1}$ $\rm (\AA)$ \\
\hline
spherical & 120 & 17.86  & 1.16 & 1.09 & 1.27 & 23.19 & 21.99 & 21.21 & 13.37\\
\vspace{3mm}
spherical & 1050 & 17.16 & 1.09 & 1.11 & 1.22 & 22.20 & 21.04 & 20.20 & 13.96\\
random & 520 (1) & 14.87  & 1.29 & 0.88 & 1.14 & 16.26 & 15.25
& 15.03 & 14.12 \\
\vspace{1.5mm}
random & 520 (2) & 16.13  & 1.09 & 0.70 & 0.76 & 18.93 & 18.36
& 17.21 & 15.91 \\
random & 1300 (1) & 14.33  & 0.69 & 1.24 & 0.85 & 16.00 & 14.85
& 14.53 & 13.71 \\
\vspace{3mm}
random & 1300 (2) & 16.27  & 0.65 & 1.85 & 1.21 & 21.01 & 19.91
& 19.05 & 12.96 \\
cylindrical & 1850 & *  & 1.33 & * & * & 17.94 & 17.12 & 16.47 & 7.16 \\
\vspace{3mm}
 & & & & & & 13.68 & 12.90 & 12.16 & 7.72\\
cyl + salt & 400 & *  & 1.29 & * & * & 17.14 & 16.45 & 15.69 & 7.60 \\
 & & & & & & 13.78 & 13.09 & 12.44 & 6.27\\
\end{tabular}
\end{center}
\end{table}

\newpage

\newpage

\newpage

\newpage

\newpage

\newpage

\newpage

\newpage

\newpage

\newpage

\newpage 

\newpage

\newpage

\newpage

\newpage

\newpage

\newpage

\newpage

\end{document}